\documentclass[a4paper,fleqn]{cas-sc}

\usepackage[authoryear,longnamesfirst]{natbib}
\usepackage{amsmath,amssymb,amsfonts}
\usepackage{algorithmic}
\usepackage{graphicx}
\usepackage{textcomp}
\usepackage{xcolor}
\usepackage{acronym}
\usepackage{caption}        
\usepackage{ragged2e}       
\usepackage[ruled,vlined]{algorithm2e}
\usepackage{array}
\usepackage{url}
\usepackage{verbatim}
\usepackage{subcaption}     
\usepackage{multirow}
\usepackage{booktabs}
\usepackage{bm}
\usepackage{enumitem}
\usepackage{orcidlink} 
\usepackage{hyperref}  
\usepackage{placeins}
\usepackage{float}
\def\BibTeX{{\rm B\kern-.05em{\sc i\kern-.025em b}\kern-.08em
    T\kern-.1667em\lower.7ex\hbox{E}\kern-.125emX}}

\acrodef{LLM}{Large Language Model}
\acrodef{VQA}{Visual Question Answering}
\acrodef{CDR}{Cross-Domain Recommendation}
\acrodef{PPCDR}{Privacy-Preserving Cross-Domain Recommendation}
\acrodef{CF-RecSys}{Collaborative Filtering-based Recommender Systems}
\acrodef{SF-UBM}{Semantic-enhanced Federated User Behavior Modeling}
\acrodef{FKD}{Fact-counter Knowledge Distillation}
\acrodef{JLT}{Johnson-Lindenstrauss Transform}
\acrodef{PLM}{Pre-trained Language Model}
\acrodef{CF}{Collaborative Filtering}
\acrodef{UBM}{User Behavior Modeling}

\def\tsc#1{\csdef{#1}{\textsc{\lowercase{#1}}\xspace}}
\tsc{WGM}
\tsc{QE}
\tsc{EP}
\tsc{PMS}
\tsc{BEC}
\tsc{DE}

\begin{document}
\let\WriteBookmarks\relax
\def\floatpagepagefraction{1}
\def\textpagefraction{.001}
\shorttitle{Federated User Behavior Modeling for LLM Recommendation}
\shortauthors{Hongyun Yang et al.}

\title [mode = title]{Federated User Behavior Modeling for Privacy-Preserving LLM Recommendation}                      
\author[1]{Hongyun Yang}
\author[1]{Shicheng Wang}
\author[2]{Xinhua Wang}
\cormark[1]
\ead{wangxinhua@sdnu.edu.cn}
\author[1]{Xinchang Zhang}
\author[3]{Hui Liu}
\author[1]{Guo Lei}
\cormark[1]
\ead{leiguo.cs@gmail.com}

\affiliation[1]{organization={School of Computer Science and Artificial Intelligence, Shandong Normal University}, country={China}}
\affiliation[2]{organization={Library, Shandong Normal University}, country={China}}
\affiliation[3]{organization={School of Computing and Artificial Intelligence, Shandong University of Finance and Economics}, country={China}}

\cortext[cor1]{Corresponding author: Xinhua Wang, Lei Guo}


\begin{abstract}
Large Language Models (LLMs) have shown great success in recommender systems. However, the limited and sparse nature of user data often restricts the LLM's ability to effectively model behavior patterns. To address this, existing studies have explored cross-domain solutions by conducting Cross-Domain Recommendation (CDR) tasks. But previous methods typically assume domains are overlapped and can be accessed readily. None of the LLM methods address the privacy-preserving issues in the CDR settings, that is, Privacy-Preserving Cross-Domain Recommendation (PPCDR). 
Conducting non-overlapping PPCDR with LLM is challenging since: 1) The inability to share user identity or behavioral data across domains impedes effective cross-domain alignment. 2) The heterogeneity of data modalities (e.g., textual vs. ID-based features) across domains complicates knowledge integration. 3) 
Fusing collaborative filtering signals from traditional recommendation models with LLMs is difficult, as they operate within distinct feature spaces.
To address the above issues, we propose SF-UBM, a Semantic-enhanced Federated User Behavior Modeling method as our solution. Specifically, to deal with \textbf{Challenge 1}, we leverage natural language as a universal bridge to connect disjoint domains via a semantic-enhanced federated architecture. Here, text-based item representations are encrypted and shared, while user-specific data remains local. To handle \textbf{Challenge 2}, we design a Fact-counter Knowledge Distillation (FKD) module to integrate domain-agnostic knowledge with domain-specific knowledge, across different data modalities.
To tackle \textbf{Challenge 3}, we project pre-learned user preferences and cross-domain item representations into the soft prompt space, aligning behavioral and semantic spaces for effective LLM learning. 
We conduct extensive experiments on three pairs of real-world domains, and the experimental results demonstrate the effectiveness of SF-UBM compared to the recent SOTA methods. 
Our code will be publicly available at: https://github.com/Nexus-Yang/SF-UBM\_master.
\end{abstract}

\begin{keywords}

Cross-Domain Recommendation \sep User Behavior Modeling  \sep Federated Learning \sep Large Language Model \sep Privacy Protection
\end{keywords}

\maketitle

\section{Introduction}
In recent years, \ac{LLM}, as a versatile form of artificial intelligence, has demonstrated remarkable success across a variety of domains, such as natural language understanding \citep{dong2019unified,liu2019multi,10729345},
information retrieval \citep{labruna2024retrieve,dai2024cocktail,dai2024bias},
and \ac{VQA} \citep{lan2023improving,guo2023images,hu2024bliva,schwenk2022okvqa},
owing to their strong capabilities in semantic understanding, pattern recognition, and logical reasoning.
For example, \cite{dong2019unified} employ a shared Transformer network and implement customized self-attention masks to govern the contextual dependencies for model predictions. \cite{dai2024cocktail} construct a corpus comprising human-authored and \ac{LLM}-generated text in diverse retrieval tasks and domains. Motivated by concerns over potential data contamination in \ac{LLM}, they further develop a temporally relevant dataset containing queries sourced from recent events. \cite{guo2023images} introduce a modular framework designed to provide \ac{LLM} prompts, thereby empowering \ac{LLM} to execute zero-shot \ac{VQA} tasks without end-to-end training.

The ability of \ac{LLM} to grasp detailed user preferences and behavior patterns has also made them a promising solution in the realm of recommender systems \citep{kim2025driven,qiu2024unveiling,lyu2023llm, xi2025rise}.
For instance, ~\cite{qiu2024unveiling} propose a plug-and-play framework that synergizes \ac{LLM} and knowledge graphs to discover user preferences, significantly enhancing the performance and interpretability of existing conversational recommender systems. ~\cite{lyu2023llm} integrate four different prompt text enriching strategies, leveraging \ac{LLM} to enhance textual data and better align with user preferences. This approach improves personalized text recommendation, significantly improving recommendation quality.
A crucial factor behind the effectiveness of the above methods is their ability to model and utilize the extensive user behavior data.
However, in real-world scenarios, user behaviors are often limited and sparse. In fact, not all users are inclined to interact with the system regularly and provide ample feedback, which significantly restricts the ability of \ac{LLM} to offer accurate recommendations~\citep{huang2019lscd, zang2022survey}.

Existing approaches enhancing \ac{LLM} recommendation that address the data-sparsity issues can be categorized as data augmentation and instruction-guided fine-tuning methods. 
For instance, LLMRec~\citep{wei2024llmrec} introduces three LLM-based graph augmentation strategies—reinforcing user-item interaction edges, enhancing item attribute modeling, and constructing user profiles—to mitigate the challenges of sparse implicit feedback and low-quality auxiliary information. TALLRec~\citep{bao2023tallrec} designs an efficient instruction-tuning framework that aligns \ac{LLM}s with recommendation tasks using limited recommendation data, thereby enabling better comprehension of user preferences. 
However, the aforementioned methods remain confined to single-domain settings and fail to fully exploit the correlations across different data domains. 
In practice, user interests are inherently multi-faceted—users typically express their preferences through positive interactions across multiple platforms or domains~\citep{chen2024survey}. 
Recent studies have also resort to cross-domain solutions by conducting \ac{CDR} tasks~\citep{guo2021gcn,guo2023disentangled,cao2023mitigating}. 
For example, ~\cite{vajjala2024cross} formulate two specialized prompt designs for \ac{CDR}. These designs aim to utilize the reasoning abilities of \ac{LLM} to mitigate data sparsity issues. ~\cite{liu2025bridge} propose a hierarchical \ac{LLM} profiling module to summarize user cross-domain preferences and a trainable adapter with contrastive regularization to adapt the \ac{CDR} task.
But the above methods typically assume domains are overlapped and need to utilize shared-data (users or items) to connect domains.
Moreover, previous methods implicitly assume that cross-domain user behavior data is readily accessible and can be aggregated for centralized training. None of the \ac{LLM} methods address the privacy-preserving issues, that is, the \ac{PPCDR} task.

Traditional approaches for \ac{PPCDR} often resort to federated learning, which enables multiple organizations or domains to collaboratively train recommendation models without directly exchanging raw user data~\citep{meihan2022fedcdr, zhang2025intelligent, chen2023winwin, zhang2023communication}. 
For instance, ~\cite{zhao2024personalized} construct personalized cross-domain preference bridges to alleviate the cold-start issue via federated meta-learning, where a common bridge is first trained through federated aggregation and subsequently refined into individualized versions via meta-network adaptation. 
But they mainly concentrate on the sequential patterns of user behaviors, with little attention given to the semantic meaning behind them. In light of this, we seek to address \ac{PPCDR} by utilizing the semantic understanding and reasoning capabilities of \ac{LLM} in \ac{UBM}, which provides us opportunities to conduct more precise recommendations.

However, addressing \ac{PPCDR} via \ac{LLM}s is a challenging task, since:
1) Due to privacy concerns, we cannot assume that user identity information can be directly shared across different domains. While previous approaches~\citep{liu2024learning} have attempted to address this issue by federated learning, they still require uploading users' personal data, such as user embeddings or user-related model parameters, to a central server, which remains a risk of exposing users' private information.
Furthermore, the majority of current methods investigate the problem of \ac{PPCDR} within overlapping scenarios, leaving the non-overlapping aspect largely unexamined. This non-overlapping condition presents considerable difficulties in terms of aligning and connecting data across various domains.
2) In previous \ac{PPCDR} methods, federated learning is typically employed to learn the public information across domains, which is then combined with the private information of the current domain to generate recommendation results. However, since these two sets of information usually come from distinct distributions, such as textual data and user ID modality, it is not feasible to directly integrate them.
3) Although \ac{LLM}s excel in user modeling and reasoning, they lack the ability to capture collaborative features among users, which is the strength of traditional recommendation models. 
But it is challenging to incorporate them to enhance \ac{LLM}s because they operate in distinct feature spaces.
Traditional models primarily capture user patterns in their behavior sequences, while \ac{LLM}s model user behaviors in the semantic space. It is crucial to develop methods to bridge the gap between these two different feature spaces.

To deal with the aforementioned challenges, this work targets at \ac{PPCDR} by proposing \ac{SF-UBM}. Unlike existing methods, our approach specifically targets the non-overlapping setting of \ac{PPCDR}, as this scenario is more common in real-world applications.
Specifically, to tackle \textbf{Challenge 1}, we resort to the universality of natural language (such as item descriptions), within a semantic-enhanced federated architecture. This approach not only fulfills privacy protection requirements but also acts as a bridge to link different non-overlapping domains.
To leverage cross-domain information, our approach treats individual domains as clients and interconnects them via item textual semantics, followed by server-side clustering to establish cross-domain semantic linkages. To protect user privacy, text-representation embeddings are shared across domains in an encrypted form, while user-associated embeddings are strictly restricted.
To handle \textbf{Challenge 2}, we integrate the domain-shared information in text-modality and the domain-specific information in ID-modality by devising a \ac{FKD} method.
In our work, the domain-agnostic information is learned by the federated network, representing the information can be shared across different domains.
The domain-specific information represents the unique characteristics to each domain, which remain localized to their respective domains to preserve privacy and relevance.
By utilizing \ac{FKD} across domains, we can convey shared domain information to local clients, which allows us to effectively transfer global knowledge while ensuring that each local client retains and enhances its distinct domain-specific characteristics. 
To deal with \textbf{Challenge 3}, we further conduct model adaptation to map the pre-learned item encoding as well as user preferences to the \ac{LLM} space via soft prompt learning as shown in Section \ref{subsec:llm mapping}.
To be more specific, during the fine-tuning stage, we create two projection layers that simultaneously convert user preferences from pre-trained recommendation models and cross-domain item representations into prompts for \ac{LLM}s. These prompts are then used as instructions and guidance for the \ac{LLM} to generate item predictions.
By doing this, we can establish a bidirectional translation bridge between discrete behavioral spaces and continuous semantic spaces, enabling effective knowledge fusion while preserving domain-specific strengths.

Our contributions are summarized as follows:
\begin{itemize}   
    \item We leverage the universality of natural language as bridges to connect disjoint domains by devising a semantic-enhanced federated learning architecture.    
    \item We solve the modality mismatch issue between domain-shared and domain-specific information by devising a \ac{FKD} method to integrate cross-modality knowledge. 
    \item We address model adaptation by projecting pre-learned recommendation model as well as cross-domain item representation to a mixed-prompt for \ac{LLM} learning. 
    \item We conduct extensive experiments on three pairs of domains in real-world datasets, and the results demonstrate the superiority of our \ac{SF-UBM} method compared with the recent SOTA methods.
\end{itemize}

\section{Related Work}
In this section, we consider federated \ac{CDR}, text and ID-based recommendation, \ac{LLM}-based behavior modeling and recommendation as our related works.

\subsection{Federated \acf{CDR}}
\ac{CDR} is the task that aims to conduct recommendations by leveraging data from multiple domains. 
A key assumption in previous methods is the ability to access information from all domains, without accounting for data privacy concerns, potentially resulting in information leakage.
Subsequently, researchers have also paid attention to the \ac{PPCDR} task.
Typical approaches have sought to address this issue by leveraging federated learning as a solution~\citep{mai2023vertical,wan2023fedpdd,chen2023win,cheng2024survey}. For example, ~\cite{wan2023fedpdd} formulate a privacy-preserving double distillation framework to tackle the challenge of sparse user overlap. This method leverages two distinct distillation processes to capture both explicit and implicit knowledge representations, combined with an offline training protocol aimed at reducing privacy risks. ~\cite{mai2023vertical} integrate graph neural networks to improve privacy-preserving user modeling. Their approach employs randomized projection for neighbor embedding aggregation and updates public gradients perturbed by ternary quantization mechanisms, effectively safeguarding user-interaction privacy while maintaining model utility.

Due to data privacy issues in real applications, there is usually no overlapping information between different domains or platforms.
Hence, recent studies have further studied \ac{PPCDR} in its non-overlapping setting~\citep{PFCR,liu2024mcrpl,FFMSR,guo2025automated,guo2023dan,guo2025semantic}. 
For instance, ~\cite{PFCR} adopt vector quantization method to analyze semantic correlations in product description texts spanning multiple domains, establishing a collaborative content representation framework within a federated learning paradigm for cross-domain knowledge transfer. ~\cite{FFMSR} directly utilize the multi-layer semantic encodings produced by \ac{PLM} within the client model to address the problem of semantic loss in ~\citep{PFCR}. 
But these two methods mainly focus on modeling the sequential patterns of user behaviors, while overlooking deep semantic information within the sentences, as well as the logical reasoning that can be drawn from such information. They do not capture the nuanced relationships and higher-level contextual understanding that are crucial for making more accurate and meaningful recommendations.

\subsection{Text and ID-based Recommendation}

Traditional recommendation method mainly focus on representing users and items by their unique IDs, based on which they can further capture the user's behavior patterns and collaborative features~\citep{yuan2019simple,tang2018personalized,kang2018self,hidasi2015session}. Typical methods like Caser~\citep{tang2018personalized} and NextItNet~\citep{yuan2019simple} employ convolutional neural networks to decode localized sequential dependencies by reinterpreting item sequences via 2D convolutional operations.
Recently, with the emergence of \ac{LLM}s, recent studies have gradually shifted towards using natural language descriptors to replace traditional IDs in user and item encodings~\citep{van2024interests,zhou2024language,guo2022reinforcement}. ~\cite{van2024interests} first generate an `ideal' course description for items based on the user's query. Then, these descriptions are converted into a search vector. By which, they can find actual courses with similar content by comparing embedding similarities. 
Another line of research~\citep{ding2021zero,mu2022id,hou2022towards,tang2025one} uses textual descriptors (e.g., item titles, descriptions) to harness richer semantic knowledge.
For example, ~\cite{hou2022towards} propose a parameter-efficient encoding architecture that combines parametric whitening transforms with a mixture of expert-enhanced adapter module to learn item representations.

But as ID and text are two kinds of modalities, they can capture different aspects of user behavior and item features. Using only one of them can only get suboptimal results.
Then, recent studies tend to integrate ID signals derived by \ac{CF} methods with semantic representations derived from textual data~\citep{cheng2024empowering,li2024enhancing,xu2024sequence,zhang2024id,guo2025semantic}. 
For instance, ~\cite{cheng2024empowering} present an end-to-end dual-stream architecture designed to model user preferences from ID and textual modalities. Their framework integrates two contrastive learning objectives to align cross-modal representations, utilizing a dual-stream encoder to dynamically model inter-modal relationships. ~\cite{li2024enhancing} adopt a decoupled training strategy that isolates ID and textual modality learning, thus mitigating the imbalance problem inherent in direct fusion approaches. 
But the above studies mainly focus on the single-domain scenario, the fusion mechanism that can integrate these two modalities in privacy-preserving \ac{CDR} tasks is largely unexplored.

\subsection{\ac{LLM}-based Behavior Modeling and Recommendation Methods}
The rise of \ac{LLM}s has significantly reshaped traditional behavior modeling and recommendation methods, particularly those that rely on behavioral ID sequences. Conventional approaches primarily concentrated on capturing user preferences through these ID sequences to perform \ac{UBM}~\citep{zhou2018atrank,dong2025prompt,zhang2024causality, wen2024learning}. For instance, ~\cite{zhou2018atrank} tackle the heterogeneity of user behaviors by projecting various behavioral types into multiple latent semantic spaces, where interactions between behaviors are modeled using self-attention mechanisms. This method allows the behavioral vectors to be applied in downstream tasks via basic attention mechanisms. In contrast, ~\cite{zhang2024causality} introduce counterfactual reasoning to evaluate the causal impacts of behavioral sequences on model outcomes, using a token-level weighting approach to adjust attention across different item tokens.

With the capabilities of \ac{LLM}s to deeply exploit latent semantic representations, \ac{LLM}-based models enable a more profound understanding of behavior, revolutionizing the way behavioral data is modeled in recommendation systems~\citep{bao2024improved,jalan2024llm}. For example, ~\cite{lee2025sealr} integrate candidate items generated by sequential models along with user behavioral data into the \ac{LLM} input space, enhancing recommendation quality. Similarly, ~\cite{jalan2024llm} propose a hybrid architecture combining \ac{LLM}s with BERT~\citep{devlin2019bert} to boost personalization in session-based social recommendation systems.
Despite these advancements, existing approaches typically focus on directly feeding semantic embeddings into \ac{LLM}s, without fully addressing the alignment of feature spaces. Furthermore, most of these methods overlook the critical issue of privacy-preservation, which remains largely unexplored in \ac{LLM}-based methods.

\section{Methodology}
\begin{figure}
  \centering
  \includegraphics[width=\textwidth]{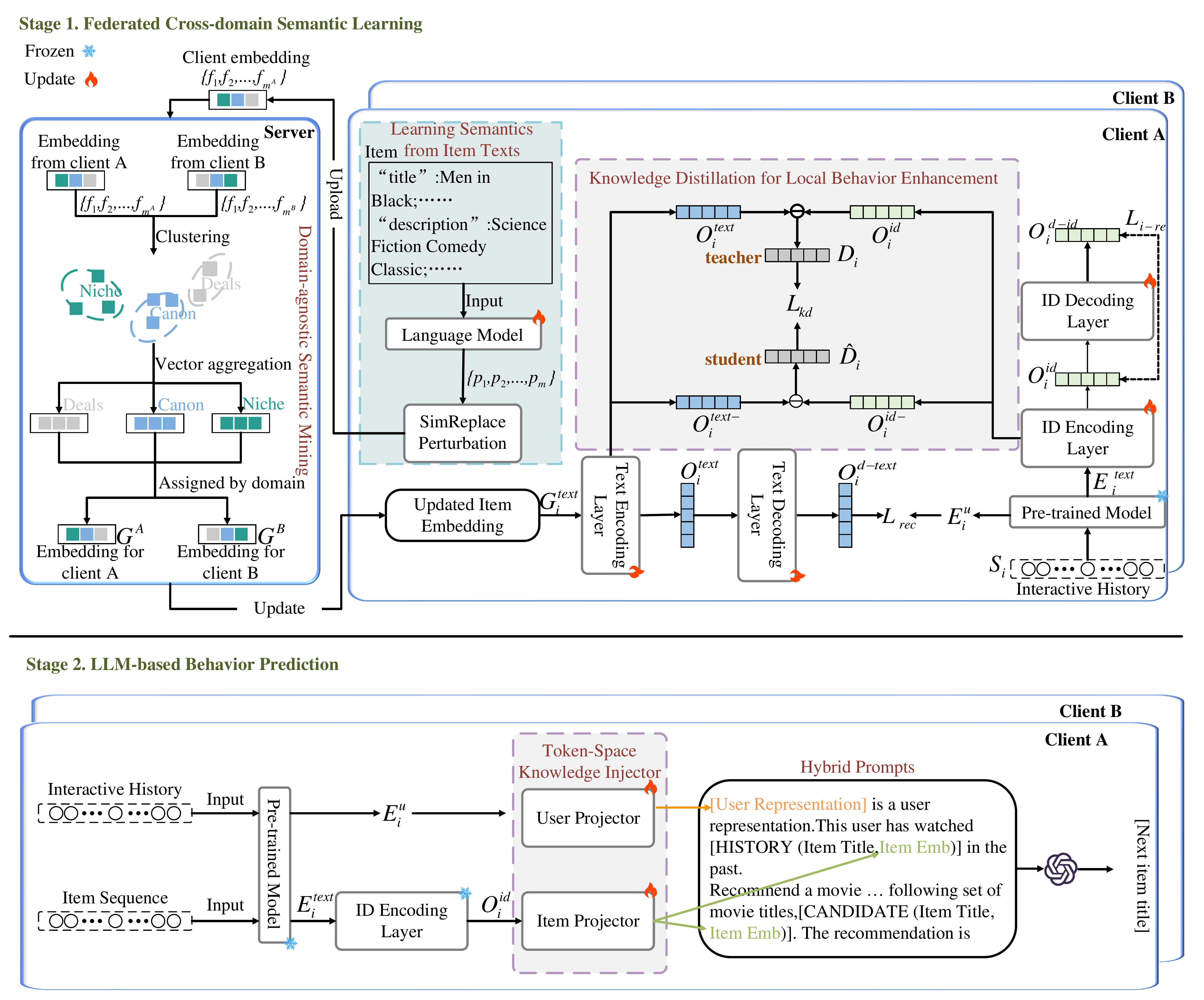}
  \caption{The system architecture of \ac{SF-UBM}, comprising two stages: 1) federated cross-domain semantic learning, where item semantics are aggregated with privacy protection and transferred to ID embeddings via \ac{FKD}; 2) \ac{LLM}-based behavior prediction, where pre-learned knowledge is projected into \ac{LLM} prompts for fine-tuning.}
  \label{fig:workflow} 
\end{figure}

In this section, we begin by defining the \ac{PPCDR} task and presenting the overall framework of our \ac{SF-UBM} solution. Subsequently, we provide a detailed explanation of how we implement \ac{SF-UBM} to achieve our objectives.

\subsection{Preliminaries}
\label{subsec:problem_definition} 
Assuming we have source domain A, target domain B, and we can obtain the sequential behaviors of the user in both domains.
Different from traditional \ac{PPCDR}, we address \ac{PPCDR} in a more challenging scenario by assuming users and items within these domains are completely disjoint. 
More formally, let $\mathcal{V}^A = \{v_1^A, v_2^A, \dots, v_{t^A}^A\}$ be the item set in domain A and $\mathcal{V}^B = \{v_1^B, v_2^B, \dots, v_{t^B}^B\}$ be the item set in domain B, where $t^A$ and $t^B$ are the number of items in A and B (suppose that A and B do not have any overlapped items). 
Then, for a given user $u_i^B \in \mathcal{U}^B$ in domain B (or $u_i^A \in \mathcal{U}^A$ in domain A), we can achieve her (or his) historical behaviors $\mathcal{S}_i^B = \{v_1^B, v_2^B, \dots, v_{m^B}^B\}$ (or $\mathcal{S}_i^A = \{v_1^A, v_2^A, \dots, v_{m^A}^A\}$), where $m^B$ and $m^A$ represent the sequence length of $\mathcal{S}_i^B$ and $\mathcal{S}_i^A$, respectively. The set of user interactions in domain A is defined as $\mathcal{S}^A$, and the set of user interactions in domain B is defined as $\mathcal{S}^B$.
In our non-overlapping scenario, both users in domain A and B are completely disjoint.
That is, we cannot directly connect domains through the shared users between them.
To maximize the generality of natural language, we assume that each element $v_j^A$ in A is associated with a textual information $\mathcal{W}_j^A = \{w_1^A, w_2^A, \dots, w_{h^A}^A\}$ ($h^A$ represents the length of the textual information for item $v_j^A$), and vice versa for items $v_j^B$ in domain B. This shared textual data enables us to connect the two disjoint domains.

In this work, each domain is regarded as a client. Our goal is to utilize both text information and unique IDs on the target domain B to accurately build a mapping function $f(\mathcal{x})$ that predicts a probability distribution over the item set $\mathcal{V}^B$. This is achieved by exploring users' historical behaviors and the knowledge transferred from the source domain A. To meet the privacy-preserving constraint, only items' text information is transferred across domains. All sensitive user-related information, such as user-item interactions, user profiles, and locally learned embeddings, remains strictly on the client device for local use.

\subsection{Overview of \ac{SF-UBM}}
\label{subsec:overview} 

\textbf{Motivation}.
To address \ac{PPCDR} in non-overlapping scenarios, we devise a federated cross-domain semantic learning framework, which simultaneously considers the text information and item IDs of user behaviors. Currently, prior research on \ac{PPCDR} primarily focuses on the ID features of user behaviors, with insufficient attention to semantic information. Intuitively, text and ID features are two modalities that can capture distinct aspects of user behavior patterns and should be treated differently.
Moreover, considering the generality of natural language across disjoint domains and the privacy-preserving constraint, we first learn items' domain-agnostic semantic features by federated learning, and then transfer them to the locally trained ID-based user behavior representations for further enhancement. The \ac{FKD} method is thereafter developed for text and ID information fusion.
To enhance the prediction of user behavior, we map the acquired semantic data into the \ac{LLM} space and leverage the \ac{LLM}'s advanced reasoning capabilities to make direct predictions based on the pre-learned cross-domain knowledge.
Additionally, given the significant distribution gap between the \ac{LLM} and the pre-learned cross-domain knowledge, we align embeddings from heterogeneous feature spaces into a unified space, adapting the domain knowledge to enhance the \ac{LLM}'s user behavior modeling capability in the target domain.

\textbf{Overall Framework}.
The overall system architecture of \ac{SF-UBM} is depicted in Fig.~\ref{fig:workflow}, which consists of two stages: federated cross-domain semantic learning, and \ac{LLM}-based behavior prediction.
\textbf{Stage 1} aims to extract more comprehensive semantic information under the privacy-preserving constraints by treating all item text descriptions within the domain as public knowledge. It begins with extracting item text semantics using SBERT~\citep{reimers2019sentence} for initial embeddings, followed by an improved encryption strategy to prevent information leakage. The encrypted embeddings are then uploaded to the server for domain-agnostic semantic mining via clustering method. To enhance local behavior modeling, a \ac{FKD} module is proposed, which combines ID-modality sequential patterns with federated text semantics, employing knowledge distillation to align multi-modal representations and reconstruction losses to prevent overly smooth embeddings.
\textbf{Stage 2} introduces pre-learned knowledge into \ac{LLM}s using two projectors: user representations from the pre-trained model and enhanced item embeddings from the ID Encoding Layer, which are mapped into the \ac{LLM}'s latent space as soft prompts. This hybrid prompt architecture combines static task instructions with these dynamic projected embeddings, allowing the \ac{LLM} to utilize cross-domain semantic connections and sequential behavior patterns for next-item prediction, optimized through token-level prediction loss.

\subsection{Stage 1: Federated Cross-domain Semantic Learning}
\label{subsec:domain federated} 
To connect distinct domains, we utilize the flexibility of natural language (served as the domain-agnostic information). In order to maintain privacy, we adopt federated cross-domain semantic learning using encrypted item texts, with user-related information being processed solely in a local manner (served as the domain-dependent information).
Thereafter, to transfer the shared domain knowledge to enhance the item's local representations, a \ac{FKD} module is further devised.

\subsubsection{Learning Semantics from Item Texts}
To extract the semantic meanings from items' description texts within specific domains, we first encode them with pre-trained large language models, and then apply encryption strategies on the learned representations for privacy-preserving. Since both Item Text Encoding and the Encryption Strategy are conducted locally within each client, we omit the superscript for brevity.

\textbf{Item Text Encoding.}
To better capture the semantic information in item description texts, we employ a sentence embedding model based on a \ac{PLM}, which generates textual semantic representations of items and systematically organizes the textual data into structured embeddings. 
In this work, the SBERT~\citep{reimers2019sentence} model is exploited. 
Specifically, for the item's text sequence $\mathcal{W}_j$, SBERT processes the text into encoded representations and applies mean-pooling to the generated token embeddings to obtain the embedding $\boldsymbol{p}_j \in \mathbb{R}^{z}$, where the value of $z$ is set to 768 in our study:

\begin{equation}
\label{deqn_ex1a1}
\boldsymbol{p}_j = \text{SBERT} ([w_1, w_2, \dots,  w_h]).
\end{equation}
This structured semantic representation facilitates subsequent privacy-aware federated operations while preserving the granularity of the item feature. 

\textbf{Encryption Strategy.}
Although only item text embeddings are uploaded to the public server, these personalized embeddings encode private behavioral patterns. An adversary could exploit inter-embedding similarities to infer sensitive user preferences underlying the local ID sequence.
To address this, we propose a replacement perturbation mechanism for encrypting embeddings. Unlike prior additive noise methods~\citep{yuan2024fellas,liang2024differentially}, which can be filtered through denoising techniques~\citep{miao2024efficient}, our approach fundamentally alters embedding positions in the feature space by selecting replacements that maximize feature-space distance while preserving task-relevant semantics.

1) Noise Perturbation. We first inject noises into the text-modality-based item embedding via the following process:

\begin{equation}
\label{deqn_ex1a2}
\boldsymbol{p}'_j= \boldsymbol{p}_j + \boldsymbol{\epsilon}_j,
\end{equation}
where $\boldsymbol{\epsilon} \sim \mathcal{N}(0, \sigma^2 \boldsymbol{I})$ is the noise introduced. $\boldsymbol{I}$ is a normal distribution with zero mean and standard deviation. $\sigma$ is used to control the standard deviation when noise is generated.

2) Semantic Preservation. Then, we perform semantics preservation through feature replacement to ensure enhanced privacy protection and reduce the semantic space distortion introduced by the added noise $\epsilon$ .
We first calculate the cosine similarity score between any two perturbed item representations $\boldsymbol{p}'_j$ and $\boldsymbol{p}'_{j'}$: 
\begin{equation}
\label{deqn_ex1a3}
sim(\boldsymbol{p}'_j, \boldsymbol{p}'_{j'}) = \frac{\boldsymbol{p}'_j\boldsymbol{p}'_{j'}}{\|\boldsymbol{p}'_j\| \|\boldsymbol{p}'_{j'}\|}.
\end{equation}

Let vector $\boldsymbol{r}_j$ represent the similarity scores between item $v_j$ and all the other items, and the Kronecker delta vector~\citep{kouw2019review} denotes influence of the item itself. Then, the similarity score vector $\boldsymbol{R}_j$ is reached:
\begin{equation}
\label{deqn_ex1a4}
\boldsymbol{R}_j = \boldsymbol{r}_j - 2 \boldsymbol{\delta}_j,
\end{equation}
where ${\boldsymbol{R}_j}$ represents the similarity scores between item $v_j$ and all the other items after excluding its own influence. Among them, $\boldsymbol{\delta}_j$ satisfies the following constraints:
\begin{equation}
\boldsymbol{\delta}_j[v_{j'}] = 
\begin{cases} 
1, & \text{if } v_{j'} = v_j, \\
0, & \text{otherwise}.
\end{cases}
\end{equation}

Subsequently, we locate the neighbor item $v_{j'}$ that has highest similarity score ($\text{max}~\boldsymbol{R}_j[j']$) to item $v_j$.
Finally, we perform similarity replacement to construct a new semantic representation for $v_j$:
\begin{equation}
\label{deqn_ex1a6}
\boldsymbol{f}_j = \boldsymbol{p}'_{j'},
\end{equation}
where $\boldsymbol{f}_j$ represents the replaced vector of item $v_j$.

The nearest neighbor constraint limits sampling to a local set of highly similar candidates, maintaining the topological structure of the feature space. This ensures enhanced privacy protection while minimizing any impact on the semantic consistency for downstream tasks, effectively striking a balance between privacy and utility.

\subsubsection{Domain-agnostic Semantic Mining on the Server}
\label{subsec:federated}
After learning the semantic information of items in each client, we then upload them to the server and mine the domain-shared information by utilizing the generality of natural language as bridges to connect disjoint domains.
Let $\mathcal{F}$ be the set of item embeddings in both clients A and B:
\begin{equation}
\mathcal{F}=\{\boldsymbol{f}_1^A, \boldsymbol{f}_2^A, \dots, \boldsymbol{f}_{t_A}^A;\boldsymbol{f}_1^B, \boldsymbol{f}_2^B, \dots, \boldsymbol{f}_{t_B}^B\},
\end{equation}
where $t_A$ and $t_B$ are the item numbers of domains A and B, respectively. We may let $ t_A + t_B = t$. For simplicity, on the server side, we omit the superscript and directly use $\boldsymbol{f}_j$ to denote the $j$-th original text embedding in the server.

\textbf{Semantic Clustering.}
To learn the shared information between domains from the semantic embeddings, we introduce K-means++~\citep{arthur2006k} as our clustering algorithm. 
It is worth noting that in our semantic clustering approach, even if some clusters contain items exclusively from a single domain, the risk of negative transfer is inherently mitigated by the nature of text embeddings. Unlike ID-based representations that are strictly domain-specific, text embeddings derived from pre-trained language models capture general semantic concepts that transcend domain boundaries. The semantic space serves as a natural bridge that enables knowledge transfer based on conceptual similarity rather than domain membership. Moreover, our similarity-based replacement mechanism further ensures that only semantically relevant cluster centroids contribute to each item's representation, effectively filtering out irrelevant domain-specific information.

Specifically, we undertake semantic clustering through the following steps:
In the initialization phase, we randomly select the first centroid from the given embeddings. We then determine the subsequent centroids based on the likelihood of selection, which is proportional to the squared distance of a data point from the nearest existing centroid. This process continues until a total of $k-1$ centroids have been chosen.
Next, using the selected $k$ centroids, we further partition the data points into $k$ clusters. This assignment involves calculating the distance between each data point and all the current centroids, assigning each point to the cluster corresponding to the nearest centroid. As a result, we obtain $k$ distinct clusters.
During the centroid updating phase, we utilize the mean aggregation method from the classic algorithm to refresh the centroids. Specifically, the new centroid for each cluster is determined by calculating the mean value of all data points that belong to that cluster.
The aforementioned process is repeated until the algorithm converges, which is assessed based on the variations in the positions of the updated centroids.

\textbf{Synchronization.}
By employing semantic clustering, we can categorize item text encodings from various domains that share similar semantics into the same semantic cluster. To facilitate the exchange of semantics across different domains, we subsequently substitute the original item text semantic encoding with the corresponding centroid vector.
The item texts processed by federated learning are represented as $\mathcal{G}^A$ and $\mathcal{G}^B$ for domain A and B, respectively, which will be returned to the clients as a form of global domain information.

\subsubsection{\acf{FKD} for Local Behavior Enhancement}
\label{subsec:modeling}
In the earlier stages, we capitalized on the flexibility of natural language to identify domain-shared information, effectively serving as bridges that link disjoint domains. To better capture the various aspects of item characteristics, we leverage the ID-modality of items within local clients to model users' sequential behaviors. Additionally, to improve this learning process, we devise the \ac{FKD} method by considering two kinds of modalities. For simplicity, we hereafter omit the superscript and conduct all operations within the same client.

\textbf{Sequential Behavior Modeling in ID-modality.}
ID-modality-based sequential recommendation methods have achieved considerable success within \ac{CF}-based models.
Typical approaches exploit Markov chains~\citep{rendle2010factorizing}, recurrent neural networks~\citep{guo2023dan,FFMSR}, and transformer-based techniques~\citep{guo2025automated} to predict users' next behaviors.
In this study, we pre-train SASRec~\citep{kang2018self} on each client for ID-based behavior modeling using cross-entropy loss, then freeze all parameters including the embedding and Transformer layers.

For a given behavior sequence $\mathcal{S}_i = (v_1, v_2, \dots, v_{m})$, each item ID is mapped to a dense embedding vector and processed through transformer blocks with multi-head self-attention.
The output provides context-aware representations with $\boldsymbol{E}^{id}_{i,j} \in \mathbb{R}^{d}$ denoting the $j$-th item encoding, where $d$ is the embedding dimension.
We use $\boldsymbol{E}^u_i \in \mathbb{R}^{m \times d}$ to represent user $u_i$'s embedding, where $m$ is the interaction sequence length.

However, ID-modality-based methods focus solely on item IDs while overlooking inherent semantic information, limiting the model's ability to capture item characteristics and user preferences. Integrating semantic information could enable a more holistic understanding and improve recommendation performance.

\textbf{\acf{FKD}.}
To leverage the domain-shared information in text-modality to enhance the local models in ID-modality, we propose a knowledge distillation method \ac{FKD} to incorporate two types of characteristics.
Intuitively, the sequence encoding in text-modality should have a similar semantic meaning to its embeddings in ID-modality, reflected by a smaller pairwise distance.
On the contrary, the preferences of different users in these two modalities should have opposite meanings.

\textit{Factual User Preferences.} 
For a given user $u_i$, and her interaction sequence $\mathcal{S}_i = (v_1, v_2, \dots, v_{m})$, we can get its text representation $\boldsymbol{E}_i^{id}$, and ID representation $\boldsymbol{G}_i^{text}$.
Although $\boldsymbol{E}_i^{id}$ and $\boldsymbol{G}_i^{text}$ represent  $\mathcal{S}_i$ in two different modalities, they denote the same user.
To promote a higher similarity between modalities, we define the factual user preference as follows:
\begin{equation}
\label{eq:factual_d}
    \boldsymbol{D}_i = \operatorname{Encoding}_{id}(\boldsymbol{E}_i^{id})-\operatorname{Encoding}_{text}(\boldsymbol{G}_i^{text}),
\end{equation}
where $\operatorname{Encoding}_{text}$ and $\operatorname{Encoding}_{id}$ are two encoders of text embedding and ID representation, one of the functions of which is to align two different embeddings. We denote the encoded information of the two modalities as $\boldsymbol{O}_i^{id}=\operatorname{Encoding}_{id}(\boldsymbol{E}_i^{id})$ and $\boldsymbol{O}_i^{text}=\operatorname{Encoding}_{text}(\boldsymbol{G}_i^{text})$, respectively.

\textit{Counterfactual User Preferences.} 
For the interactions that are not interacted by user (random selection is applied), they reflect the user preferences in the counterfactual world. In this context, for a behavior sequence originating from the counterfactual world, its representations in different modalities will still express similar semantic information. Therefore, the distance between these two representations should remain relatively close.
Such distance is defined as follows:
\begin{equation}
\label{eq:counterfactual_d}
    \boldsymbol{\hat{D}}_i =\operatorname{Encoding}_{id}(\boldsymbol{E}_i^{id-})-\operatorname{Encoding}_{text}(\boldsymbol{G}_i^{text-}),
\end{equation}
where $\boldsymbol{E}_i^{id-}$ and $\boldsymbol{G}_i^{text-}$ denote user preferences in the counterfactual sequence of the text and ID modality, respectively.

\textit{Knowledge Distillation.}
For the same user, her interest and preference in both the factual world and the counterfactual world, i.e., $\boldsymbol{D}_i$ and $\boldsymbol{\hat{D}}_i$, should be consistent. In order to transmit information from the counterfactual world to the factual world, we use their difference between the two modalities as a mediator. For user $u_i$, we define the student mediator as the element-wise difference between the embeddings in ID and text modalities from the factual world, as shown in Eq. (\ref{eq:factual_d}).
For the teacher mediator, we define it as the element-wise difference between two modalities in the counterfactual world, as shown in Eq. (\ref{eq:counterfactual_d}). 
Based on the teacher and student mediators, the distillation loss can be defined as:
\begin{equation}
\mathcal{L}_{kd} = \frac{1}{|U|} \sum |\boldsymbol{D}_i - \boldsymbol{\hat{D}}_i|^2,
\end{equation}
where $|U|$ represents the number of users in the target client.

Moreover, to prevent overly smooth representations under extreme conditions, we introduce a reconstruction decoder along with a reconstruction loss function. We learn more effective item representations by calculating the following reconstructive losses:

\begin{equation}
    \mathcal{L}_{t-re}=\underset{\boldsymbol{S}_i \in \mathcal{S}}{\mathbb{E}}[\text{MSE}(\boldsymbol{G}_i^{text},\operatorname{Decoding}_{text}(\boldsymbol{O}_i^{text})],
\end{equation}
\begin{equation}
    \mathcal{L}_{i-re}=\underset{\boldsymbol{S}_i \in \mathcal{S}}{\mathbb{E}}[\text{MSE}(\boldsymbol{E}_i^{id},\operatorname{Decoding}_{id}(\boldsymbol{O}_i^{id})],
\end{equation}
where $\operatorname{Decoding}_{id}$ serves as the decoder for the ID modality, and $\operatorname{Decoding}_{text}$ is defined for the text modality. Correspondingly, we denote their decoded vectors as $\boldsymbol{O}_i^{d-id}=\operatorname{Decoding}_{id}(\boldsymbol{O}_i^{id})$ and $\boldsymbol{O}_i^{d-text}$ is a sequential text embedding generated by the local text encoding model.

\subsubsection{Optimization Objective}
Beyond performing \ac{FKD} between the collaborative knowledge derived from user-item interactions and the textual knowledge extracted from the federated text information, we further introduce a recommendation loss. This loss not only explicitly integrates collaborative knowledge into the model, but also imparts task-specific guidance for the recommendation task. Specifically, the definition of the recommendation loss is given as follows~\citep{kang2018self}:
\begin{equation}
\begin{aligned}
\mathcal{L}_{\mathrm{rec}} = - \sum_{\boldsymbol{S}_{i} \in \mathcal{S}}[\log(\sigma(s(\boldsymbol{E}^{u-1}_i, \boldsymbol{O}_i^{d-id}))) \\
+\log(1 -\sigma(s( \boldsymbol{E}^{u-1}_i, \boldsymbol{O}_i^{d-id-})))],
\end{aligned}
\end{equation}
where $\boldsymbol{E}^{u-1}_i$ represents the user behavior representation obtained by the pre-trained model from the interaction sequence of users after removing the last interaction item, and $s(a,b)$ is a dot product between $a$ and $b$.

To this end, the overall learning objective is as follows:
\begin{equation}
\begin{aligned}
\mathcal{L} = \mathcal{L}_{kd}+\alpha\mathcal{L}_{t-re}+\beta\mathcal{L}_{i-re}+\mathcal{L}_{rec},
\end{aligned}
\end{equation}
where $\alpha$ and $\beta$ are the hyper-parameters that balance the reconstruction loss and prevent the generation of over-smoothed representations.

\subsubsection{Privacy Preservation Analysis.}

Conventional federated learning keeps data local, yet recent work shows that shared gradients or parameters can still leak user information. In non-overlapping cross-domain recommendation, this risk remains.

We propose a two-layer lightweight privacy mechanism. First, for each item \(v_i\), a Sentence-BERT encoder maps its textual side information (title, category, etc.) into a normalized embedding \(\tilde{\mathbf{e}}_i^{\text{text}}\). Only this semantic vector – containing no user interaction data – is considered for upload, so the server cannot infer any user's behavior history. Second, before upload, Gaussian noise is added:
\begin{equation}
\hat{\mathbf{e}}_i = \tilde{\mathbf{e}}_i + \boldsymbol{\epsilon}_i,\quad \boldsymbol{\epsilon}_i \sim \mathcal{N}(\mathbf{0},\sigma^2\mathbf{I}).
\end{equation}
Instead of sending \(\hat{\mathbf{e}}_i\) directly, the client uploads \(\mathbf{z}_i = \hat{\mathbf{e}}_{\pi(i)}\), where \(\pi(i)\) is the index of the nearest neighbor (excluding self) in the perturbed set under cosine similarity. This breaks the direct item–vector mapping.

Regarding privacy guarantee, the adversary observes \(\mathbf{z}_i\) but not \(\tilde{\mathbf{e}}_i\). Two sources of uncertainty apply: noise limits inversion precision (Fisher information \(\propto 1/\sigma^2\)), and the substitution \(\pi(i)\) randomizes item–vector correspondence. In a dense semantic space, the probability of correctly re-identifying the original item from \(\{\mathbf{z}_i\}\) is at most \(1/k\), where \(k\) is the average neighborhood size. Hence item-level privacy is provably enhanced.

The main advantage of this mechanism is that it avoids heavy encryption, incurring only one nearest-neighbor search and noise generation per item. Experiments (Section~\ref{privacy-preserving}) show that it resists inversion attacks while keeping recommendation degradation within an acceptable range.

\subsection{Stage 2: Prior Knowledge-enhanced LLM Behavior Prediction}
\label{subsec:llm mapping}
Although large models possess enhanced logical reasoning abilities, relying solely on single-domain information still leads to suboptimal results.
Thus, to leverage the pre-learned knowledge to further enhance \ac{LLM}s, we 
apply two projectors that project the cross-domain embeddings in the ID Embedding Layer and the \ac{CF} features in the Pre-trained Sequential Model into the latent space in \ac{LLM}. In this work, the OPT-6.7B~\citep{zhang2022opt} is exploited. 
The architecture of Stage 2 is shown in Fig.~\ref{fig:workflow}.

\subsubsection{Token-Space Knowledge Injector}
From Stage 1, we can obtain two kinds of prior knowledge:

The \textbf{user representation} $\boldsymbol{E}^u_i \in \mathbb{R}^{m \times d}$ captured by pre-trained traditional recommendation models (SASRec is adopted in this work) can be defined as:
\begin{equation}
\begin{aligned}
\boldsymbol{M}_u=\operatorname{Projector}_u(\boldsymbol{E}^u_i ),
\end{aligned}
\end{equation}
where $\operatorname{Projector}_u$ represents the special mapping projector aligning the user representation in the prompt space.
$\boldsymbol{M}_u \in \mathbb{R}^{token}$ denotes the user soft prompts to be injected to \ac{LLM}.
We pre-train SASRec with the goal of modeling historical interaction behaviors to capture the latent preference patterns inherent in user activities. 
This process allows us to generate personalized recommendations by identifying relationships between users and items from the \ac{CF} view, which is the core strength of traditional models.

The \textbf{enhanced item embedding} $\boldsymbol{E}_i^{id}\in \mathbb{R}^{d'}$ in the ID Encoding Layer that integrates both domain-agnostic and domain-specific information at item-level is formulated as:
\begin{equation}
\begin{aligned}
\boldsymbol{M}_i=\operatorname{Projector}_{i}(\boldsymbol{E}_i^{id}),
\end{aligned}
\end{equation}
where $\operatorname{Projector}_i$ represents the mapping function that aligns the item representation within the prompt space, 
$\boldsymbol{M}_i \in \mathbb{R}^{token}$ denotes the item soft prompts to be injected to \ac{LLM}.

The ID Encoding Layer plays a critical role by merging fundamentally different modalities of items, i.e., text and ID-modality, into unified, dense vector representations. This integration preserves ID signals while simultaneously infusing transferable domain semantic knowledge, which is crucial to effectively addressing downstream tasks.

\subsubsection{Hybrid Prompts for \ac{LLM} Prediction}
To further leverage the pre-learned prior knowledge and enhance behavior modeling in \ac{LLM}, we employ a hybrid prompting architecture that combines static instructions with dynamically projected knowledge embeddings from pre-learned models.

Specifically, the input prompt comprises two components: a static textual template serving as fixed task directives, and the projected knowledge embeddings $\boldsymbol{M}_i$ and $\boldsymbol{M}_u$ serving as the soft prompts to maintain the pre-learned knowledge. We obtain the hybrid prompt $\mathcal{P}(\boldsymbol{M}_i,\boldsymbol{M}_u)$.

\textbf{Loss Function.} With the model parameters in the Pre-trained Sequential Model and ID Encoding Layer fixed, we proceed to learn the mapping functions via the following loss function:
\begin{equation}
\begin{aligned}
\mathcal{L}_{CE} = - \sum \log p ( \boldsymbol{y}_l \mid \boldsymbol{y}_{<l},\mathcal{P}),
\end{aligned}
\end{equation}
where $p (\cdot)$ denotes the probability that the model predicts the $l$-th token (denoted as $\boldsymbol{y}_l$), given the preceding tokens $\boldsymbol{y}_{<l}$. $l$ represents the length of the token sequence.
Our learning objective of Stage 2 is to finetune a \ac{LLM} by optimizing the above loss function to enable it to precisely generate the next user behavior.

\section{Experimental Setup}
\label{sec:setup}
This section presents the key research questions explored in our experiments and provides a detailed explanation of the datasets, evaluation protocols, and baseline models used for model evaluation.

\subsection{Research Questions}
\label{subsec:rqs}
We rigorously evaluate our \ac{SF-UBM} method through the following four research questions:
\begin{itemize}
    \item \textbf{RQ1} How does our method compare to SOTA recommendation approaches on \ac{PPCDR}?
    \item \textbf{RQ2} How do the core components of \ac{SF-UBM} impact the overall recommendation performance?
    \item \textbf{RQ3} How do the key hyper-parameters influence the effectiveness of \ac{SF-UBM}?
    \item \textbf{RQ4} What impact does the level of privacy protection have on the performance of recommendations?
\end{itemize}

\subsection{Datasets}
\label{Dataset}

\begin{table}[t]
\centering
\caption{The Statistics of Our Datasets.}
\label{tab:dataset}
\small
\setlength{\tabcolsep}{8pt} 
\begin{tabular}{@{}llccc@{}}
\toprule
 & \textbf{Dataset} & \textbf{Users} & \textbf{Items} & \textbf{Avg\_Lens} \\
\midrule
\multirow{3}{*}{\textbf{Target Domain}} 
 & Health & 4,088 & 3,593 & 3.42 \\
 & Books & 6905 & 41,598 & 8.22 \\
 & Food & 12,689 & 32,228 & 7.48 \\
\cmidrule(lr){1-5} 

\multirow{3}{*}{\textbf{Source Domain}} 
 & Beauty & 8,499 & 5,828 & 6.71 \\
 & Movielens & 9,554 & 9,102 & 9.99 \\
 & Kitchen & 8,936 & 39,273 & 7.61 \\

\bottomrule
\end{tabular}
\end{table}

For comprehensive evaluations, we adopt two real-world datasets Amazon \footnote{https://nijianmo.github.io/amazon/index.html} and Movielens~\citep{harper2015movielens} as our data source, and conduct extensive experiments on three pairs of domains within them, i.e., ``Food-Kitchen", ``Books-Movielens", and ``Health-Beauty", to evaluate \ac{SF-UBM}.
All datasets consist of comprehensive textual information, including items' ``title" and ``description". 
In our datasets, ``Food-Kitchen" and ``Health-Beauty" are both from the same platform, Amazon.
``Books-Movielens" is a cross-platform dataset, where Books is from Amazon, while the other domain is from Movielens.
This is to evaluate the applicability of our method in cross-platform datasets.
Moreover, to satisfy the non-overlapping characteristic, all the users and items are kept disjoint in different domains. 

In order to make these datasets suitable for our evaluation protocols, we process them as follows.
We remove items with user ratings below 3 from the Health-Beauty dataset, and exclude sequences with fewer than 3 interactions from the Beauty dataset to validate the cold-start scenario. For the ``Food-Kitchen" and ``Books-Movielens" datasets, we refer to the dataset processing method in \cite{liao2023llara}. First, we eliminate the sequences with fewer than 5 interactions and items with ratings below 3.  
Then, to construct a representative user subset, we employ a stratified sampling approach based on user interaction frequency. Specifically, we partition the user population into multiple strata according to their total interaction counts, using intervals such as 5-10 and 10-15, and draw samples from each stratum. For each selected user, we retain their most recent 10 interaction records, or all available interactions if fewer than 10 exist. 
This method preserves authentic user behavior patterns while mitigating sampling bias across different user activity levels.
The statistical information of our datasets is presented in Table \ref{tab:dataset}.

\renewcommand{\arraystretch}{1.5}
\newcolumntype{C}{>{\centering\arraybackslash}p{1cm}}
\setlength{\tabcolsep}{5pt}

\begin{table*}[ht]
\centering
\caption{Comparison results on \textit{Health-Beauty}, \textit{Food-Kitchen}, and \textit{Books-Movielens}.}
\label{tab:comparison}
\resizebox{\textwidth}{!}{%
\renewcommand{\arraystretch}{1.3}
\begin{tabular}{@{}ll ccc ccc ccc@{}}
\toprule
\multirow{2}{*}{\textbf{Category}} & \multirow{2}{*}{\textbf{Method}} & \multicolumn{3}{c}{\textbf{Health-Beauty}} & \multicolumn{3}{c}{\textbf{Food-Kitchen}} & \multicolumn{3}{c}{\textbf{Books-Movielens}} \\
\cmidrule(lr){3-5} \cmidrule(lr){6-8} \cmidrule(lr){9-11}
 & & Valid Ratio & Hit@1 & Imp.\% & Valid Ratio & Hit@1 & Imp.\% & Valid Ratio & Hit@1 & Imp.\% \\
\midrule
\multirow{2}{*}{\textbf{ID-only}} 
& SASRec & 100\% & 0.3889 & 6.56 & 100\% & 0.2685 & 36.39 & 100\% & 0.0617 & 370.01 \\
& GRU4Rec & 100\% & \underline{0.4072} & 1.77 & 100\% & 0.0518 & 606.95 & 100\% & 0.0676 & 328.99 \\
\midrule
\multirow{2}{*}{\textbf{CDRec}} 
& PTUPCDR & 100\% & 0.0557 & 643.98 & 100\% & 0.0641 & 471.29 & 100\% & 0.0587 & 394.04 \\
& RecGURU & 100\% & 0.1763 & 135.05 & 100\% & 0.1562 & 134.44 & 100\% & 0.1049 & 176.45 \\
\midrule
\multirow{3}{*}{\textbf{ID-Text}} 
& UniSRec & 100\% & 0.1025 & 304.29 & 100\% & 0.1188 & 208.25 & 100\% & 0.1274 & 127.63 \\
& PFCR & 100\% & 0.2287 & 81.20 & 100\% & 0.1277 & 186.77 & 100\% & 0.0703 & 312.52 \\
& FFMSR & 100\% & 0.3889 & 6.56 & 100\% & 0.2806 & 30.51 & 100\% & 0.0833 & 248.14 \\
\midrule
\multirow{4}{*}{\textbf{LLM-Rec}} 
& LLM-Only & 85.53\% & 0.0220 & 1783.64 & 74.95\% & 0.0049 & 7374.47 & 90.91\% & 0.0052 & 5476.92 \\
& TALLRec & 97.63\% & 0.3350 & 23.70 & 99.35\% & \underline{0.3660} & 0.05 & 95.46\% & 0.1922 & 50.88 \\
& MLP-LLM & 89.66\% & 0.3475 & 19.25 & 99.11\% & 0.3106 & 17.90 & 97.45\% & \underline{0.2086} & 39.02 \\
& A-LLMRec & 97.48\% & 0.3735 & 10.95 & 97.89\% & 0.2920 & 25.41 & 92.80\% & 0.1757 & 65.05 \\
\midrule
\textbf{OURS} & \textbf{SF-UBM} & \textbf{98.19\%} & \textbf{0.4144} & \textbf{-} & \textbf{99.49\%} & \textbf{0.3662} & \textbf{-} & \textbf{99.15\%} & \textbf{0.2900} & \textbf{-} \\
\bottomrule
\end{tabular}%
}
\end{table*}

\subsection{Evaluation Protocols}
To systematically evaluate the performance of our method in next-behavior prediction tasks, we organize all the user interaction behaviors in chronological order and employ a leave-one-out strategy to partition user interaction sequences.
That is, utilizing the last interaction as the test data, the second-to-last as the validation set, and the remaining interactions as the training data.
This splitting approach ensures that subsequent interactions do not appear in the training data, thereby preventing any potential information leakage. 
During model training, the system uses all historical interactions before the target item, excluding the target itself to avoid data leakage.

To ensure our results are comparable with the latest SOTA recommendation methods with LLMs, we adopt the evaluation metrics utilized in A-LLMRec~\citep{kim2024large} and LLARA~\citep{liao2023llara}. Specifically, we utilize the Hit Ratio @1 metric to assess our experimental outcomes, which indicates that a successful prediction is recognized only when the ground-truth item is ranked first. Additionally, we evaluate model performance using the Valid metric, which represents the proportion of valid and correctly formatted outputs. The term Imp.\% in our results denotes the relative improvement percentage in comparison to baselines.

\subsection{Baselines}
\label{Baseline}
To demonstrate the effectiveness of our method in addressing non-overlapping \ac{PPCDR}, we compare \ac{SF-UBM} with the following four types of baselines:

1) \textbf{ID-only recommendation methods:}
  This kind of method only utilizes ID to denote items.
\begin{itemize}
    \item \textbf{SASRec~\citep{kang2018self}:} This is a self-attention-based approach that encodes users' historical interactions to produce personalized item recommendations. This is a strong baseline for sequential recommendation, achieving excellent results across multiple datasets.
    \item \textbf{GRU4Rec~\citep{hidasi2015session}:} This is a pioneering session-based recommendation method that introduces gated recurring units to capture temporal dynamics in user behavior sequences. This method is designed specifically for short-lived sessions.
\end{itemize}
  
2) \textbf{\acf{CDR} methods:} In this category, the methods do not need any overlapping information to bridge or align domains.
\begin{itemize}
    \item \textbf{PTUPCDR~\citep{zhu2022personalized}:} This is a meta-learning-based method that employs personalized bridge networks to align user representations across domains and transfer preference knowledge through adaptive meta-transfer functions.
    \item \textbf{RecGURU~\citep{li2022recguru}:} This is an adversarial learning-based framework that facilitates \ac{CDR} by utilizing generalized user representations.
\end{itemize}
  
Note that we do not compare with the overlapping \ac{CDR} methods, as they need overlapped users or an identical number of training samples in both source and target domains, which is not the case for our datasets.
  
3) \textbf{ID-Text recommendation methods:}
Methods in this category simultaneously utilize ID and Text to represent items.
\begin{itemize}
    \item \textbf{UniSRec~\citep{hou2022towards}:} This method leverages item text to create generic textual representations of items. For a thorough comparison with our approach, we trained UniSRec from scratch, using its hybrid version that combines ID embeddings with text embeddings.
    \item \textbf{PFCR~\citep{PFCR}:} This is a SOTA method for non-overlapping \ac{PPCDR}. It utilizes vector quantization to create shared cross-domain embeddings, enabling efficient information transfer between domains.
    \item \textbf{FFMSR~\citep{FFMSR}:} This is another recent SOTA method on our task. This approach employs multi-layer semantic encoding and a dynamically designed gating mechanism combined with a fast Fourier transform-based filtering layer to eliminate redundant noise within the sequences.
\end{itemize}
Both PFCR and FFMSR are recent SOTA methods specifically designed for non-overlapping PPCDR scenarios, and are evaluated under the same setting as our \ac{SF-UBM}.

4) \textbf{\ac{LLM}-based recommendation methods:}
This kind of method utilizes \ac{LLM} as the recommendation model.
\begin{itemize}
    \item \textbf{\ac{LLM}-Only:} This method leverages the open-source OPT~\citep{zhang2022opt} as the recommendation engine by applying task-specific recommendation prompts, where the OPT model with 6.7B parameters is adopted for all the following \ac{LLM}-based recommendation methods.
    \item \textbf{TALLRec~\citep{bao2023tallrec}:} This method is a prompt-based learning approach for recommendation tasks that relies exclusively on text. It utilizes LoRA to fine-tune the \ac{LLM}. The approach involves providing the model with the user's interaction history and a candidate item, after which the model generates a response indicating whether the user is likely to like the item.
    \item \textbf{A-LLMRec \citep{kim2024large}:} This is an efficient \ac{LLM}-based recommender system that achieves robust cross-scenario recommendation through a dual-stage alignment mechanism to infuse \ac{CF} knowledge into \ac{LLM}.
    \item \textbf{MLP-\ac{LLM}:} This is a variant of \ac{SF-UBM} that employs a simple MLP for representation mapping, which directly feeds embeddings from pre-trained models and \ac{LLM}-generated representations into the \ac{LLM}'s prompts. To ensure consistency, we adopt the same prompt format as in \ac{SF-UBM}.
\end{itemize}

\subsection{Implementation Details}
We implement the \ac{SF-UBM} model using the PyTorch framework. All experiments are conducted on an NVIDIA RTX 3090 GPU, with the exception of the TALLRec~\citep{bao2023tallrec} experiments, which are performed on an NVIDIA RTX 5090 GPU. The \ac{CF} model used in \ac{SF-UBM} is SASRec, and OPT-6.7B~\citep{zhang2022opt} is adopted as the backbone of the \ac{LLM}-based methods. For fairness of model performance comparison, we also use OPT-6.7B as the backbone for other \ac{LLM} models. 
In Stage 1, we use the SASRec model as the pre-trained model and fix the dimension of items in all datasets and the dimension of the model's Embedding to 50. The $\alpha$ and $\beta$ in the overall objective loss are set to 0.5 and 0.2, respectively.
In the encryption strategy, $\sigma$ for noise generation is set to 0.1, and $\sigma$ for the ``Health-Beauty" dataset is set to 0.05. In the federated cross-domain semantic fusion module, considering the differences in data volume during the experiment, the number of centroids used for clustering on the server side is set to 90 for the ``Health-Beauty" dataset, 100 for the ``Food-Kitchen" dataset, and 70 for the ``Books-Movielens" dataset. In the \ac{FKD} module, the batch size is set to 32, and the number of training epochs is set to 10. 
The maximum number of iterations for the clustering algorithm is set to 300, with the default $tol$ value of 0.0001.
For the \ac{LLM} mapping stage, we fix the data dimension to 128 in the mapping training of hybrid embeddings and ID sequence embeddings, and the two mapping functions use different activation functions, with the number of epochs set to 5. During the training phase, we use the Adam optimizer with a learning rate of 0.0001. 
For the hyper-parameters in the baselines, we follow their original settings in the publications, and fine-tune them in specific datasets.
\begin{table}[t]
\centering
\caption{Hyper-parameter settings of \ac{SF-UBM}.}
\label{tab:hyperparams-2}
\small
\setlength{\tabcolsep}{4pt}
\begin{tabular}{@{}ccccccc@{}} 
\toprule
Dataset & 
\begin{tabular}[c]{@{}c@{}}\textit{lr}\end{tabular} & 
\begin{tabular}[c]{@{}c@{}}Emb. dim\\ (pre-train)\end{tabular} & 
\begin{tabular}[c]{@{}c@{}}Emb. dim\\ (\ac{LLM})\end{tabular} & 
$\alpha$ & 
$\beta$ & 
$k$ \\
\midrule
Health-Beauty & 0.0001 & 50 & 128 & 0.5 & 0.2 & 90 \\
Food-Kitchen  & 0.0001 & 50 & 128 & 0.5 & 0.2 & 100 \\
Books-Movielens & 0.0001 & 50 & 128 & 0.5 & 0.2 & 70 \\
\bottomrule
\end{tabular}
\end{table}

\section{Experimental Results (RQ1)}
The comparison results of our method against all baselines on three domain pairs are presented in Table \ref{tab:comparison}, from which we can draw the following conclusions:
1) Our proposed \ac{SF-UBM} framework significantly outperforms other SOTA methods under the same constraints, highlighting its effectiveness in utilizing auxiliary textual information for cross-domain knowledge transfer and improved user modeling. This result also demonstrates the effectiveness of our \ac{LLM}-based solution in addressing \ac{PPCDR}. 
2) \ac{SF-UBM} surpasses ID-based approaches (e.g., SASRec, GRU4Rec) by incorporating semantic signals as supplementary features. This design allows for the extraction of latent semantic knowledge, generating hybrid embeddings that align more closely with user preferences.
3) In cross-domain scenarios, \ac{SF-UBM} outperforms competing cross-domain methods (e.g., RecGURU, PTUPCDR). The framework uniquely balances information privacy preservation and knowledge transfer under non-overlapping constraints. By relying solely on text similarity for semantic knowledge propagation, it achieves more accurate user preference extraction, making it particularly advantageous for non-overlapping \ac{CDR}s.
4) Compared to other ID-Text hybrid methods (i.e., UniSRec, PFCR, FFMSR), \ac{SF-UBM} demonstrates superior performance through a novel distillation mechanism that effectively infuses textual knowledge into ID sequences. This enables more accurate user behavior modeling using auxiliary information.
5) When compared to \ac{LLM}-based recommendation models (e.g., \ac{LLM}-Only, TALLRec), \ac{SF-UBM} shows clear advantages. While models like \ac{LLM}-Only and TALLRec rely solely on natural language inputs without multi-modal integration, and approaches such as MLP-\ac{LLM} and A-LLMRec perform simplistic cross-modal mappings without fully exploring inter-modal relationships, our distillation strategy proves far more effective in capturing complex modality interactions.

\section{Model Analysis}
\label{PERFORMANCE ANALYSIS}
\subsection{Ablation Study (RQ2)}

\begin{table*}[ht]
\centering
\caption{Ablation studies on \textit{Health-Beauty}, \textit{Food-Kitchen}, and \textit{Books-Movielens}.}
\label{tab:Ablation}
\resizebox{\textwidth}{!}{%
\renewcommand{\arraystretch}{1.3}
\begin{tabular}{@{}l ccc ccc ccc@{}}
\toprule
\multirow{2}{*}{\textbf{Method}} & \multicolumn{3}{c}{\textbf{Health-Beauty}} & \multicolumn{3}{c}{\textbf{Food-Kitchen}} & \multicolumn{3}{c}{\textbf{Books-Movielens}} \\
\cmidrule(lr){2-4} \cmidrule(lr){5-7} \cmidrule(lr){8-10}
& Valid & Hit@1 & Imp.\% & Valid & Hit@1 & Imp.\% & Valid & Hit@1 & Imp.\% \\
\midrule
Base & 85.53\% & 0.0220 & 1783.64 & 74.95\% & 0.0049 & 7373.47 & 90.91\% & 0.0052 & 5476.92 \\
Base + Text\_only & 89.66\% & 0.2675 & 35.45 & 99.11\% & 0.3227 & 13.48 & 97.45\% & 0.2435 & 19.09 \\
Base + ID\_only & 98.14\% & 0.3654 & 11.82 & 99.03\% & 0.3069 & 19.32 & 98.75\% & 0.2137 & 23.63 \\
Base + KD & 98.17\% & 0.4008 & 3.39 & 99.33\% & 0.3129 & 17.03 & 98.89\% & 0.2539 & 14.21 \\
Base + ac{FKD} & 98.17\% & 0.4075 & 1.69 & 99.40\% & 0.3357 & 9.09 & 98.94\% & 0.2569 & 12.88 \\
\midrule
\textbf{SF-UBM (Ours)} & \textbf{98.19\%} & \textbf{0.4144} & \textbf{-} & \textbf{99.49\%} & \textbf{0.3662} & \textbf{-} & \textbf{99.15\%} & \textbf{0.2900} & \textbf{-} \\
\bottomrule
\end{tabular}%
}
\end{table*}

To systematically assess the contribution of individual components in our framework, we further perform an ablation study by comparing \ac{SF-UBM} with its four degraded variants:
\begin{itemize}
    \item Base + Text\_only: This variant uses only textual information, removing the pre-trained model along with federated learning modules and all modules processing clustered encoding.
    \item Base + ID\_only: This variant removes textual information, retaining only the ID encoding obtained from pre-training by eliminating the \ac{FKD} module and all subsequent text encoding modules. These first two variants are designed to demonstrate differences in the user behavior information contained across modalities (as shown in Section~\ref{subsec:modeling}).
    \item Base + KD: This experiment performs knowledge distillation between ID and text modalities locally without leveraging auxiliary textual information from other domains (as described in Section~\ref{subsec:domain federated}). This assesses the importance of cross-domain federated learning for knowledge transfer.
    \item Base + \ac{FKD}: This variant removes the module aligning hybrid embeddings with ID sequence embeddings (as in Section~\ref{subsec:llm mapping}). This verifies the importance of mapping different embeddings to a unified feature space and employing concise fixed prompts for \ac{LLM}-based recommendation.
\end{itemize}

From the experimental results reported in Table~\ref{tab:Ablation}, we can achieve the following conclusions:
1) \ac{SF-UBM} achieves better performance than its variants (Base + Text\_only and Base + ID\_only) across all metrics. This demonstrates that while different modalities across domains contain distinct feature knowledge, \ac{FKD} effectively filters out redundant information and extracts more precise representations for user modeling. Meanwhile, the significant performance drop on the Health-Beauty dataset indicates that when interaction sequences are too short, even textual information provides limited benefit.
2) \ac{SF-UBM} performs better than \ac{SF-UBM} (Base + KD) in all performance metrics. This proves the effectiveness of federated learning in bridging non-overlapping domains under privacy-preserving constraints at the semantic level.
3) \ac{SF-UBM} outperforms \ac{SF-UBM} (Base + \ac{FKD}) across all performance metrics. This confirms the critical role of the mapping function in aligning embeddings to a unified prompt feature space.

\subsection{Hyper-parameter Analysis (RQ3)}
\begin{figure*}[t]
\centering
\begin{subfigure}{0.32\textwidth}
    \centering
    \includegraphics[width=\linewidth]{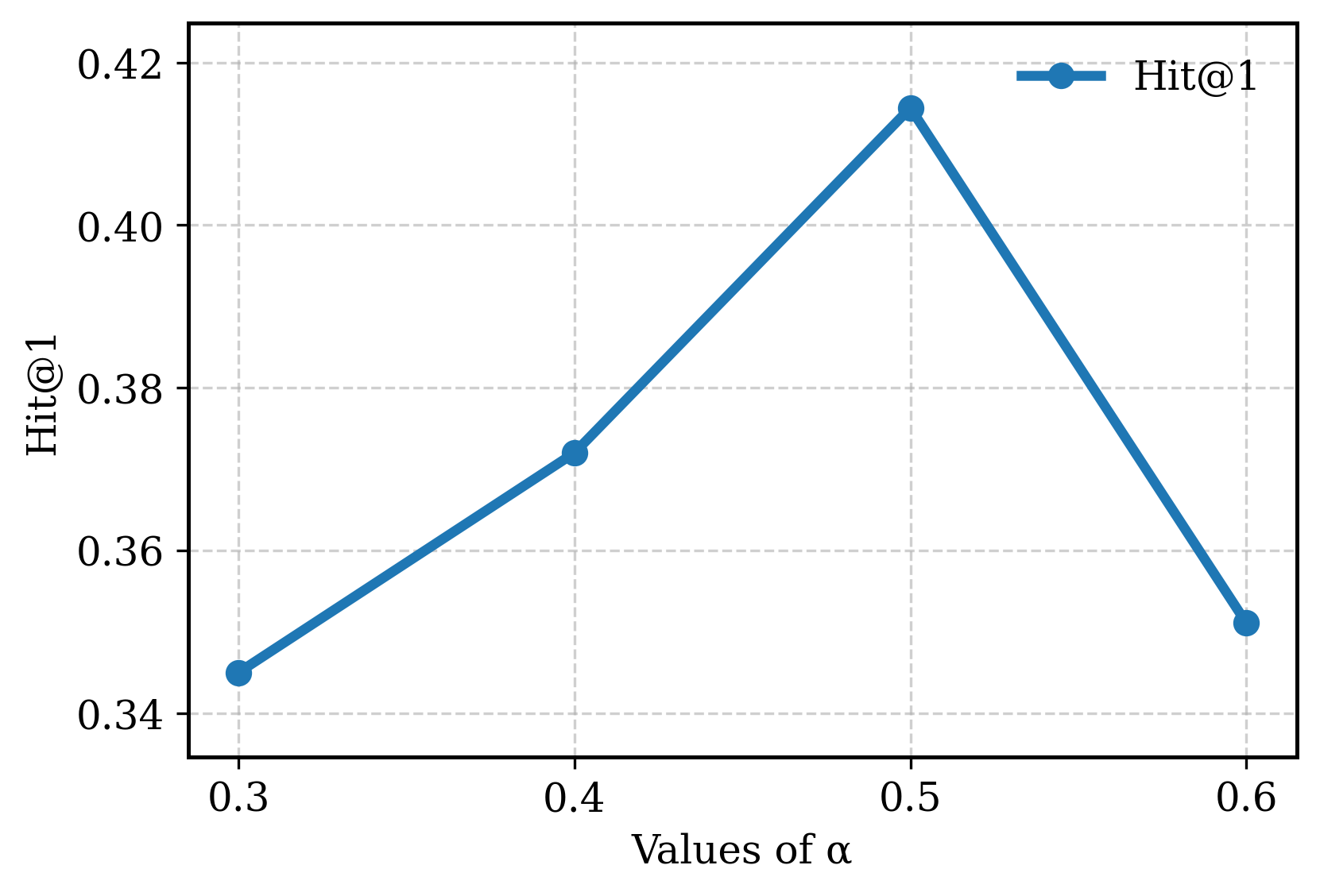}
    \caption{Health-Beauty under different $\alpha$}
    \label{fig:q_health}
\end{subfigure}
\hfill
\begin{subfigure}{0.32\textwidth}
    \centering
    \includegraphics[width=\linewidth]{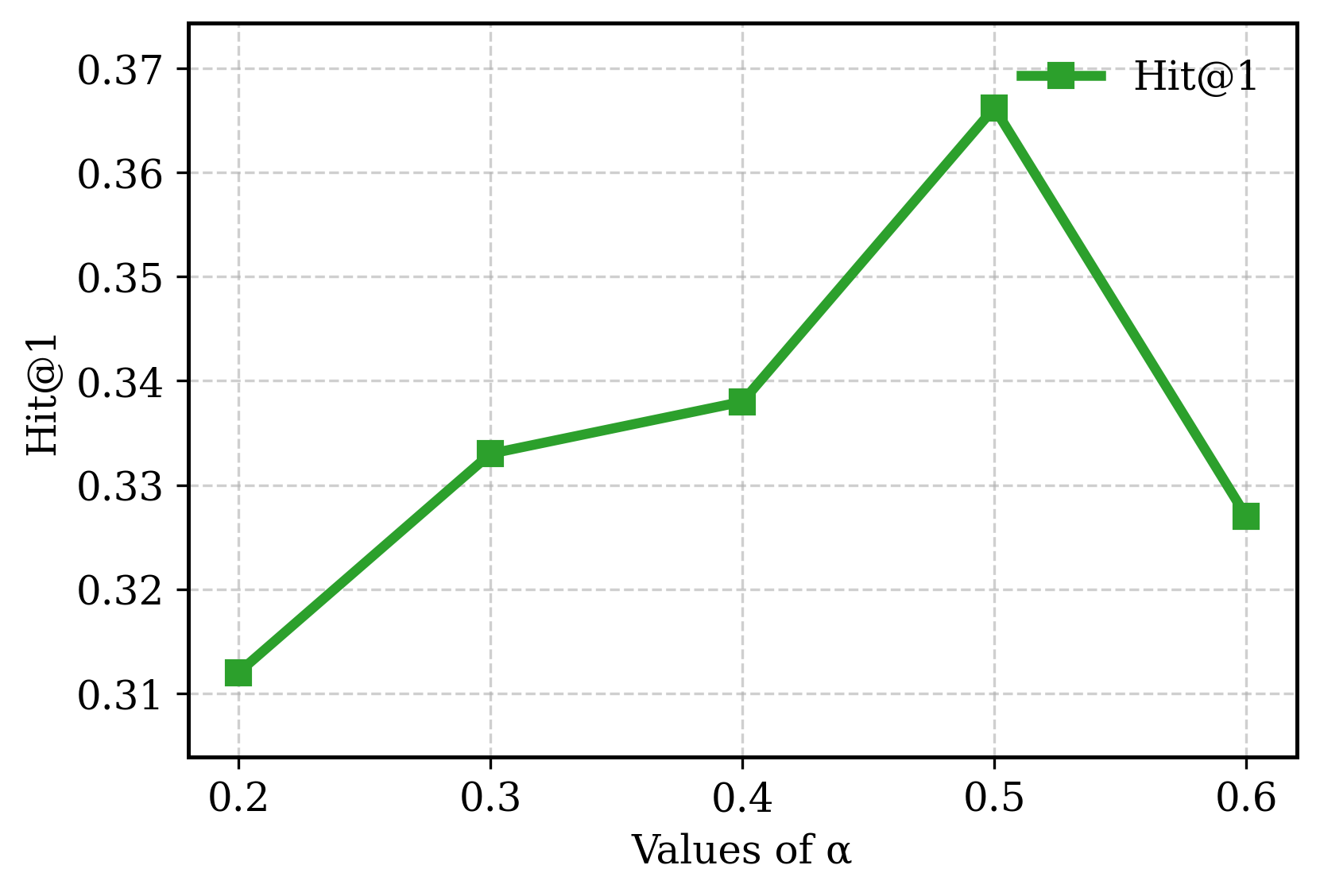}
    \caption{Food-Kitchen under different $\alpha$}
    \label{fig:q_food}
\end{subfigure}
\hfill
\begin{subfigure}{0.32\textwidth}
    \centering
    \includegraphics[width=\linewidth]{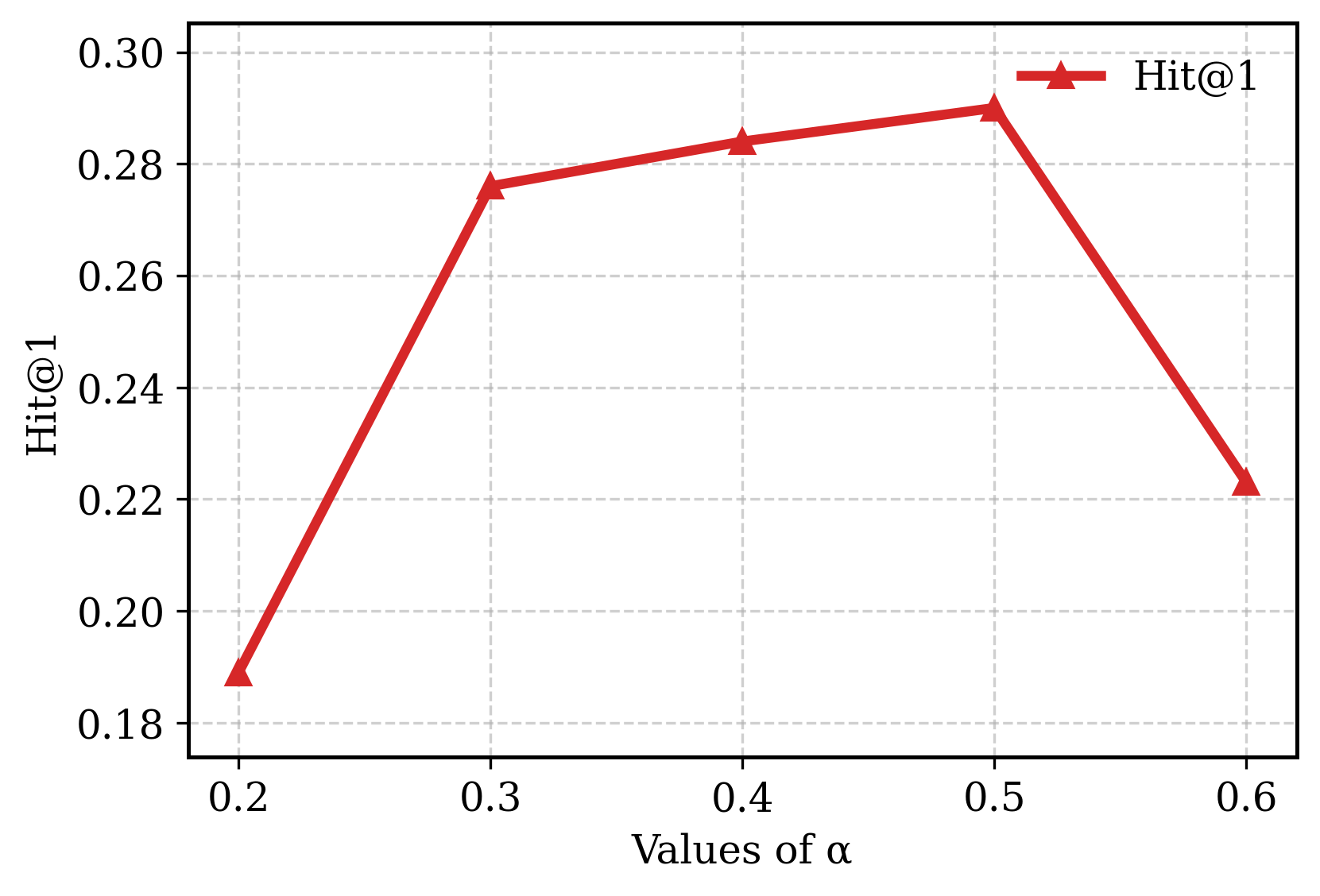}
    \caption{Books-Movielens under different $\alpha$}
    \label{fig:q_books}
\end{subfigure}
\caption{Overall performance under different $\alpha$}
\label{fig:alpha_analysis}
\end{figure*}

\begin{figure*}[t]
  \centering
  \begin{subfigure}{0.32\textwidth}
    \centering
    \includegraphics[width=\linewidth]{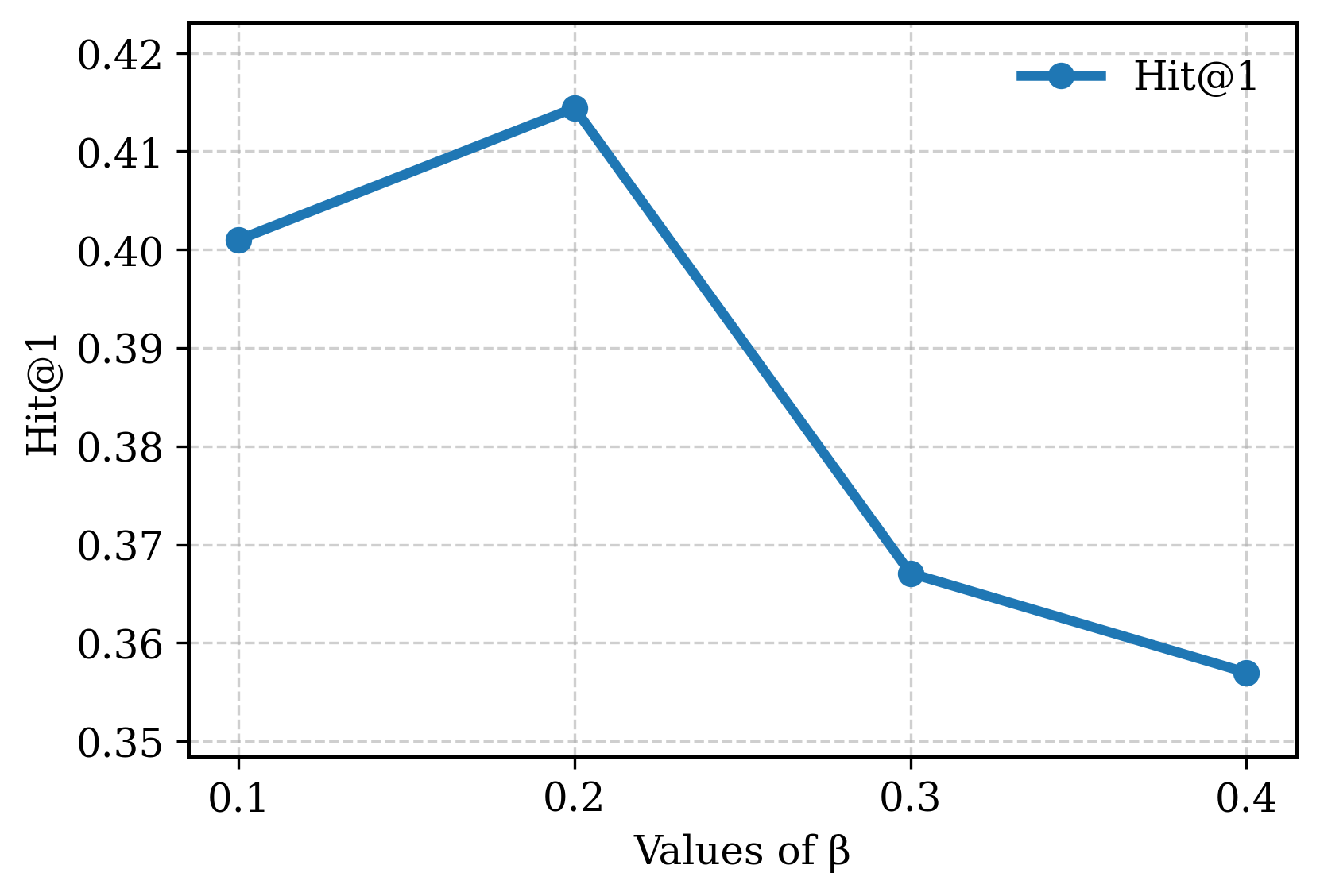}
    \caption{Health-Beauty under different $\beta$}
    \label{fig:beta_health}
  \end{subfigure}
  \hfill
  \begin{subfigure}{0.32\textwidth}
    \centering
    \includegraphics[width=\linewidth]{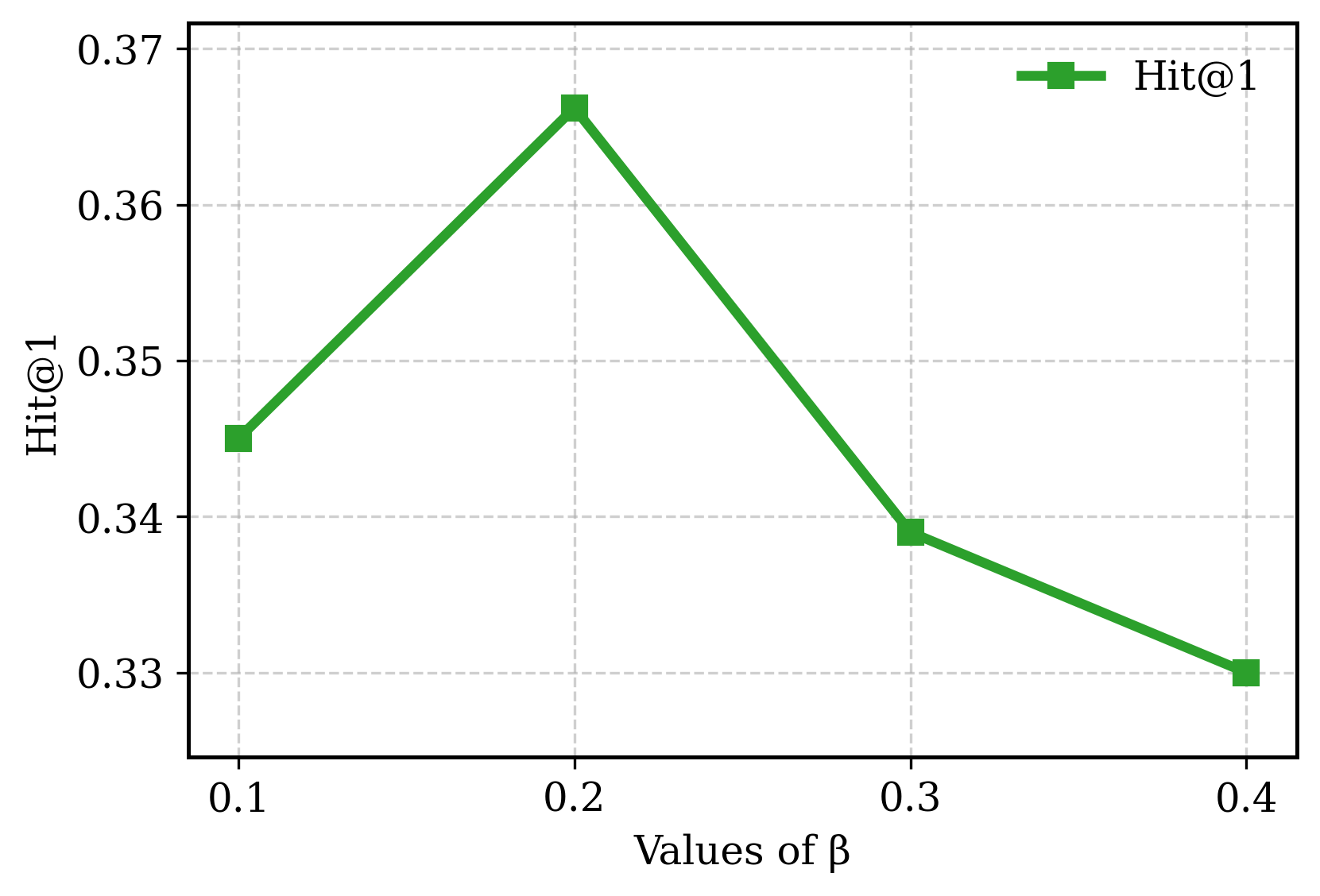}
    \caption{Food-Kitchen under different $\beta$}
    \label{fig:beta_food}
  \end{subfigure}
  \hfill
  \begin{subfigure}{0.32\textwidth}
    \centering
    \includegraphics[width=\linewidth]{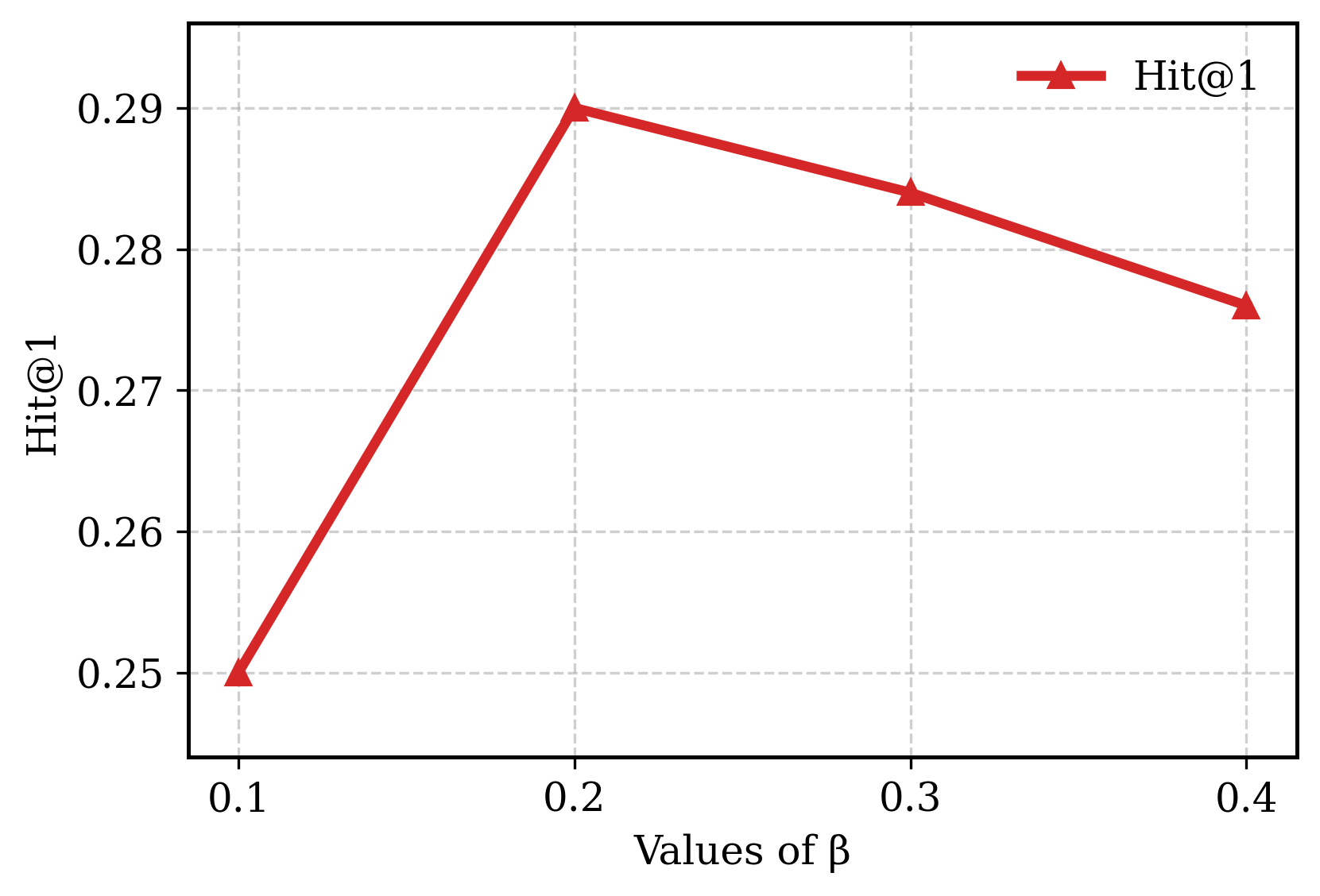}
    \caption{Books-Movielens under different $\beta$}
    \label{fig:beta_books}
  \end{subfigure}
  \caption{Overall performance under different $\beta$}
  \label{fig:beta_results}
\end{figure*}

We conduct systematic experiments on six critical hyper-parameters: cluster count $k$, learning rate, embedding dimensions, and loss weight ratios, to investigate their impacts on \ac{SF-UBM} performance.
Fig.~\ref{fig:alpha_analysis} and Fig.~\ref{fig:beta_results} illustrate how different values of $\alpha$ and $\beta$ affect \ac{SF-UBM} performance. As presented in Fig.~\ref{fig:alpha_analysis} (a-c), all three dataset pairs achieve optimal performance when $\alpha = 0.5$. Similarly, Fig.~\ref{fig:beta_results} (a-c) shows that the best results are obtained when $\beta = 0.2$.

\begin{figure*}[t]
\centering
\begin{subfigure}{0.32\textwidth}
    \centering
    \includegraphics[width=\linewidth]{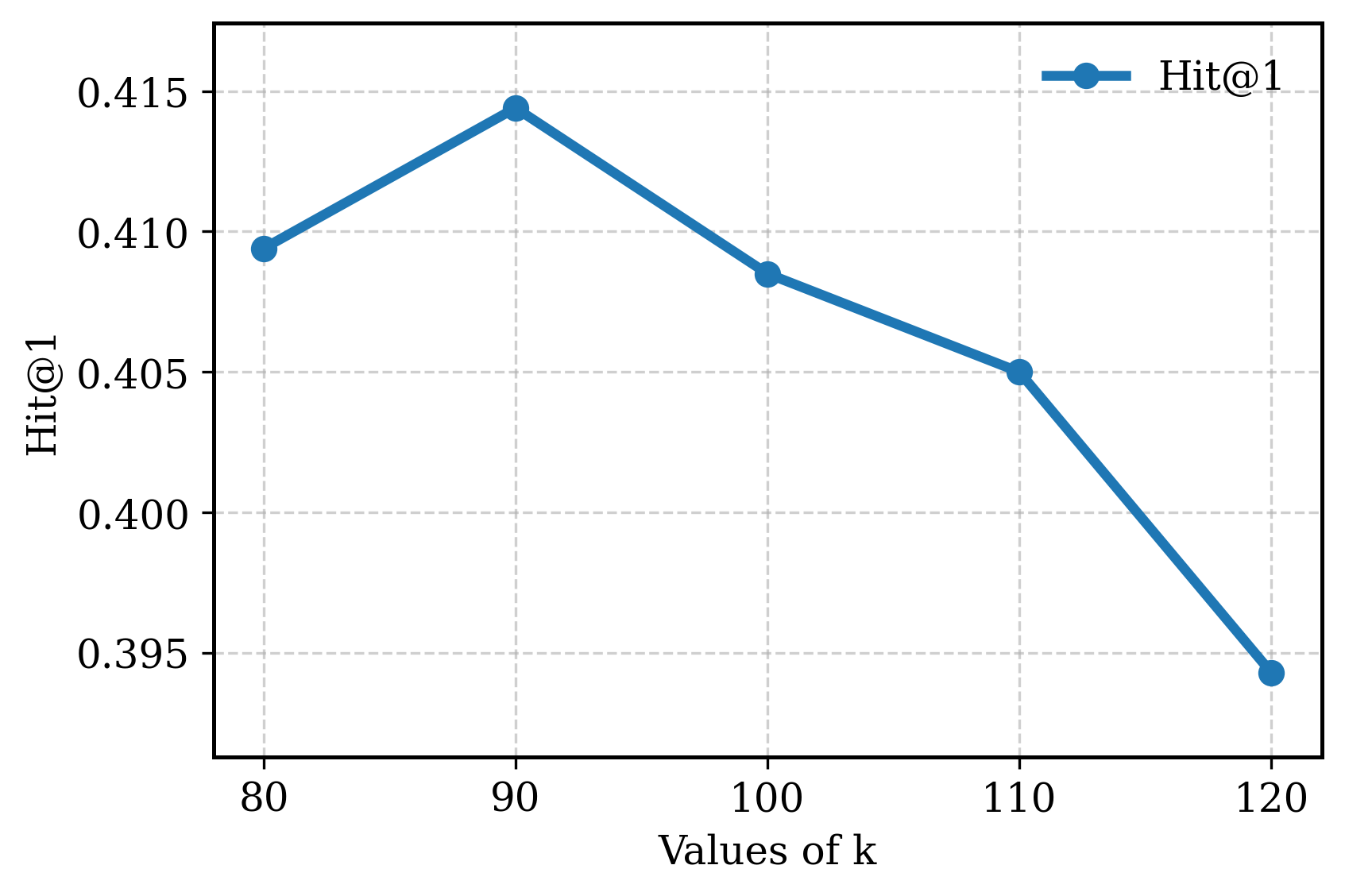}
    \caption{Health-Beauty under different $k$}
    \label{fig:k_health}
\end{subfigure}
\hfill
\begin{subfigure}{0.32\textwidth}
    \centering
    \includegraphics[width=\linewidth]{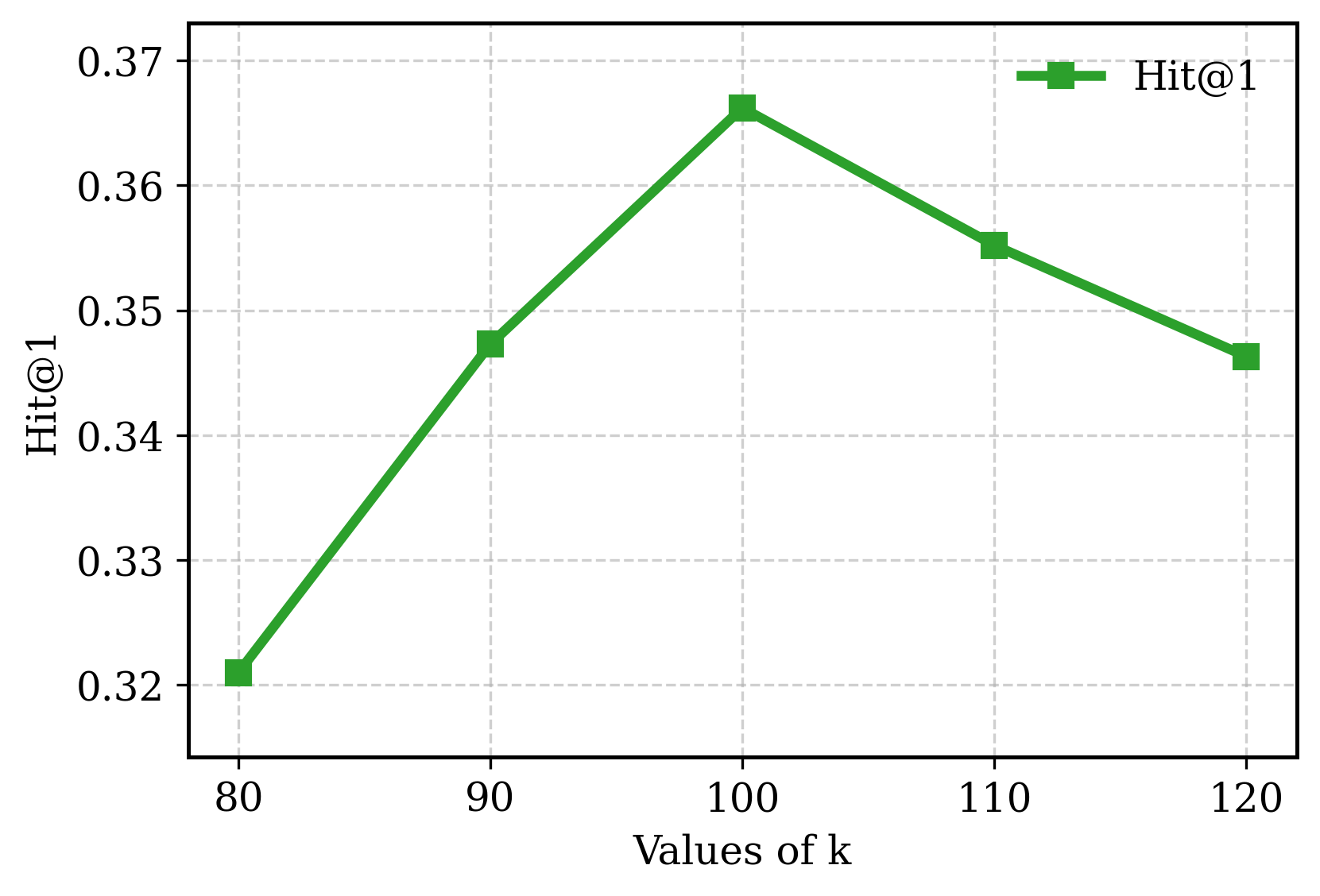}
    \caption{Food-Kitchen under different $k$}
    \label{fig:k_food}
\end{subfigure}
\hfill
\begin{subfigure}{0.32\textwidth}
    \centering
    \includegraphics[width=\linewidth]{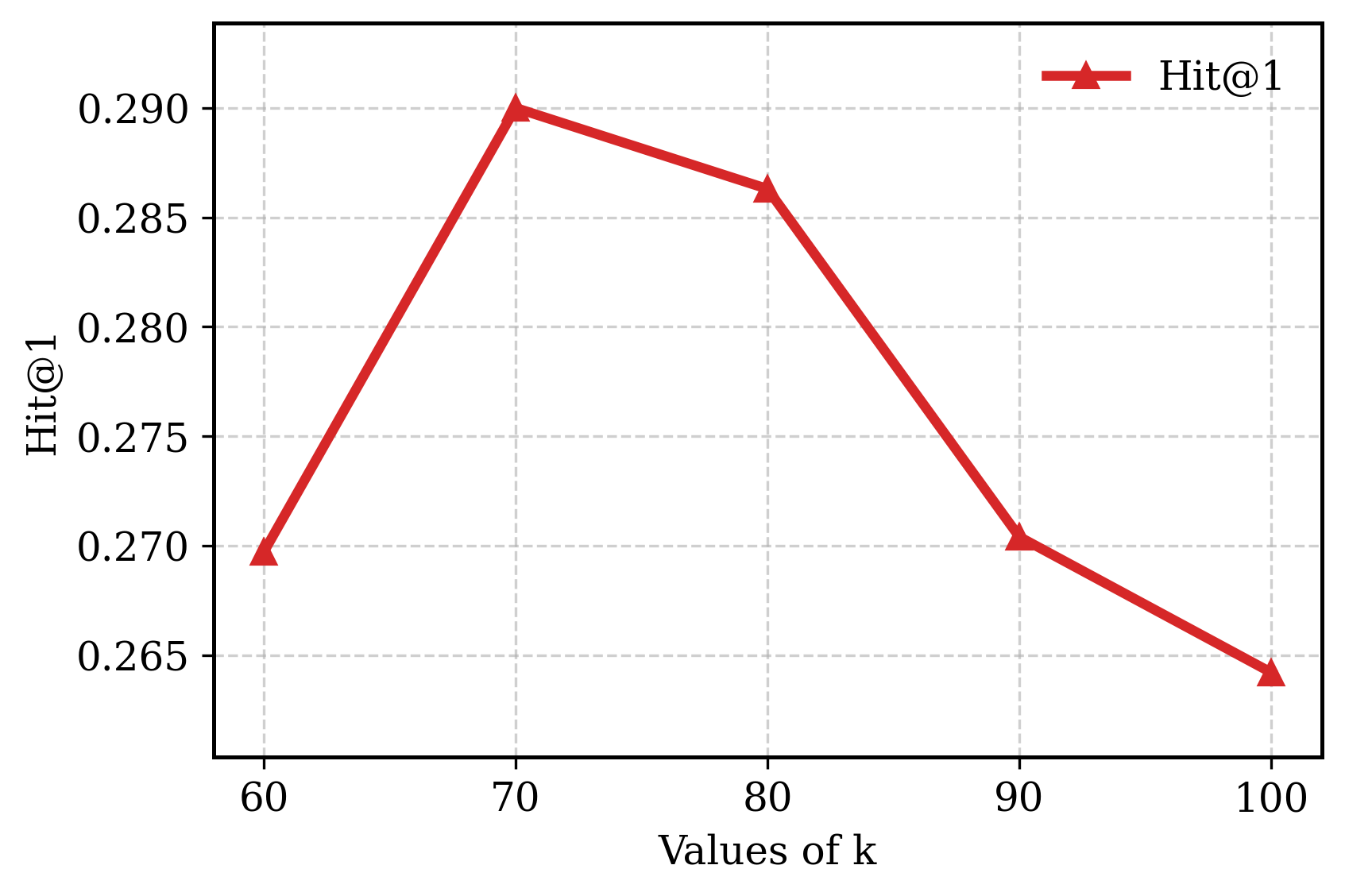}
    \caption{Books-Movielens under different $k$}
    \label{fig:k_books}
\end{subfigure}
\caption{Overall performance under different k}
\label{fig:k_analysis}
\end{figure*}

The federated clustering parameter $k$ significantly influences the effectiveness of cluster encoding. As demonstrated in Fig.~\ref{fig:k_analysis}, optimal performance is observed within the range $70 < k < 100$. Performance degrades when $k$ deviates from this interval, suggesting the existence of domain-specific optimal cluster densities. The remaining parameters (learning rate, embedding dimensions, and loss weights) exhibit varying impacts across datasets, with optimal configurations being dataset-dependent.
Beyond these observations, we report the performance-optimized hyper-parameter combinations for each dataset in Table~\ref{tab:hyperparams-2}, revealing notable variations in ideal embedding dimensions between domains with sparse versus dense interaction patterns.

\subsection{Impact of the Perturbation Level on Recommendation Performance (RQ4)}
\label{privacy-preserving}

\begin{table}[t]
\centering
\caption{Perturbation effectiveness under different noise levels.}
\label{tab:privacy-similarity}
\small
\setlength{\tabcolsep}{4pt}
\begin{tabular}{@{}ccccccc@{}}
\toprule
$\sigma$ & Health & Beauty & Food & Kitchen & Books & Movielens \\
\midrule
0.5   & 0.3263 & 0.3924 & 0.3956 & 0.3596 & 0.3584 & 0.4131 \\
0.1   & 0.5428 & 0.6354 & 0.6431 & 0.6032 & 0.5732 & 0.6410 \\
0.05  & 0.5778 & 0.6830 & 0.7018 & 0.6200 & 0.6245 & 0.6980 \\
0.005 & 0.8237 & 0.9097 & 0.8692 & 0.8309 & 0.8921 & 0.8951 \\
\bottomrule
\end{tabular}
\end{table}

In our privacy preservation mechanism, we add Gaussian noise to the item embeddings to 
obfuscate attackers and prevent them from inferring the original item textual features and semantic information from the uploaded data. Additionally, we employ a similar item substitution strategy to further enhance the protection of sensitive item information. In our approach, the parameter $\sigma$ represents the standard deviation of the added Gaussian noise, controlling the perturbation strength. A larger value of $\sigma$ indicates greater noise addition, resulting in lower similarity between perturbed and original embeddings, less preserved original semantic information, and stronger privacy protection, but may significantly impact recommendation performance. Conversely, a smaller value of $\sigma$  indicates less noise addition, allowing perturbed embeddings to maintain higher similarity with original embeddings, preserving more semantic information to maintain recommendation accuracy, but with relatively weaker privacy protection strength.

To verify the impact of our privacy defense method on the recommendation performance, we vary the hyper-parameter $\sigma$ in rage of $[0.005 - 0.5]$, and measure the change of the item embedding by calculating the similarity between item embeddings before and after adding noise.
The experimental results are shown in the Table~\ref{tab:privacy-similarity}, from which we can achieve the following conclusions: 
1) The level of perturbation is negatively correlated with recommendation performance. As the noise level $\sigma$ increases from $0.005$ to $0.5$, the cosine similarity between original and perturbed embeddings decreases from $0.82 - 0.91$ to $0.33 - 0.41$, demonstrating that increasing noise effectively obfuscates original semantic information and significantly enhances privacy protection strength. 
2) Our perturbation strategy can effectively prevent information leakage. At our noise configurations ($\sigma = 0.05$ or $\sigma = 0.1$), the item similarity before and after adding the perturbation is reduced to $0.54 - 0.70$, with approximately $30\% - 45\%$ of original information obfuscated. This ensures that our method can effectively prevent the server from inferring original features.
3) Our method have stable protection effects across various datasets. From Table \ref{tab:privacy-similarity}, we also observe that different datasets display comparable perturbation effects at the same noise levels, with similarity values varying by less than (0.1). This suggests that our method can be readily generalized across various data distributions.

\subsection{Train/Inference Speed.}
Table~\ref{tab:privacy-similarity} reports the training time, inference time, and Hit@1 of A-LLMRec and SF-UBM on two datasets. 
Training costs are considerably dataset-dependent, with Books demanding substantially more time than Health. 
Overall, the two methods remain comparable in training efficiency: SF-UBM is marginally slower on Health, yet slightly faster on Books. 
Inference time per batch is identical on the same dataset, indicating that SF-UBM introduces no extra computational overhead. 
As for recommendation accuracy, SF-UBM consistently yields a higher Hit@1, and the margin is notably larger on Books than on Health. 
Hence, under the examined conditions, SF-UBM preserves superior recommendation accuracy with comparable training and inference cost, and its advantage becomes more evident on the more time-consuming Books dataset.
\begin{table}[t]
\centering
\caption{Perturbation effectiveness under different noise levels.}
\label{tab:privacy-similarity}
\small
\setlength{\tabcolsep}{4pt}
\begin{tabular}{llccc}
\toprule
Method & Dataset & Train time (min) & Inference time  & Hit@1 \\
\midrule
\multirow{2}{*}{A-LLMRec}     & Health & 343.33 & 1.98 & 0.3735 \\
                                       & Books  & 743.5  & 2.31 & 0.1757 \\
\midrule
\multirow{2}{*}{SF-UBM} & Health & 388.58 & 1.98 & 0.4144 \\
                                       & Books  & 732.5  & 2.31 & 0.2900 \\
\bottomrule
\end{tabular}
\end{table}

\section{Conclusion}
In this paper, we presented \ac{SF-UBM}, a semantic-enhanced federated framework for \ac{PPCDR} across non-overlapping domains. It leverages natural language as a modality-agnostic bridge and introduces a \ac{FKD} module to integrate domain-specific and domain-agnostic knowledge, aligning behavioral and semantic spaces through soft-prompt tuning of a pre-trained \ac{LLM}. Evaluations on three domain pairs confirm its superiority over state-of-the-art baselines. We further supplied empirical privacy–utility analysis and communication cost breakdowns to provide a more practical view of the framework.

Despite these strengths, several limitations remain. The current perturbation strategy lacks formal privacy guarantees, and the client-side computational footprint introduced by components such as SBERT and SASRec may impede deployment on resource-constrained devices. The evaluation focuses on a single ranking metric, and certain design choices, such as the clustering algorithm, are sensitive to hyper-parameters. In future work, we intend to explore perturbation mechanisms with provable privacy bounds (e.g., local differential privacy), develop lightweight client-side alternatives, extend the evaluation to multiple recommendation metrics, investigate robust clustering techniques, and incorporate multi-modal data for stronger cross-domain alignment. We will also refine the construction of counterfactual preferences to strengthen theoretical grounding.

\appendix
\section{My Appendix}
Appendix sections are coded under \verb+\appendix+.

\verb+\printcredits+ command is used after appendix sections to list 
author credit taxonomy contribution roles tagged using \verb+\credit+ 
in frontmatter.

\printcredits

\bibliographystyle{cas-model2-names}

\bibliography{cas-refs}

\end{document}